\documentclass[fleqn, 12pt]{article}
\usepackage{natbib} 
\usepackage{amssymb} 
\usepackage{amsthm} 

\usepackage{placeins}  
\usepackage{float}
\usepackage{comment}
\usepackage{fnpct}
\usepackage{mathtools}

\def\clap#1{\hbox to 0pt{\hss#1\hss}}

\usepackage{dcolumn}  
\usepackage{changepage}
\usepackage{ragged2e} 
\usepackage[multiple, bottom]{footmisc}

\usepackage[colorlinks=true, allcolors=blue]{hyperref}
\usepackage{xurl}
\usepackage{bm}
\usepackage{amsfonts}
\usepackage{amsmath}
\usepackage{amssymb}
\usepackage{nccmath}
\usepackage[english]{babel}
\usepackage{accents}
\usepackage{url}
\usepackage{alphalph}
\usepackage{setspace} 
\usepackage{graphicx}
\usepackage[top=32.5truemm,bottom=32.5truemm,left=28truemm,right=28truemm]{geometry}

\newtheorem{Proposition}{Proposition}
\newcommand{\be}{\begin{equation}}
\newcommand{\ee}{\end{equation}}
\newcommand{\bi}{\begin{itemize}}
\newcommand{\ei}{\end{itemize}}
\newcommand{\ben}{\begin{enumerate}}
\newcommand{\een}{\end{enumerate}}
\newcommand{\bdm}{\begin{displaymath}}
\newcommand{\edm}{\end{displaymath}}
\newcommand{\pu}{\underline{\pi}}
\newcommand{\po}{\Pi}

\usepackage{etoolbox}
\patchcmd{\pprintMaketitle}{\footnotesize\itshape\elsaddress\par\vskip36pt}{\footnotesize\itshape\elsaddress\par\parbox[b][36pt]{\linewidth}{\vfill\hfill\textnormal{\today}\hfill\null\vfill}}{}{}%

\begin{document}


\begin{center}
    {\Large {\bf A Global Minimum Tax for Large Firms Only: \\ Implications for Tax Competition}\renewcommand*{\thefootnote}{\fnsymbol{footnote}}\footnote[1]{Financial support from the German Research Foundation through CRC TRR 190 (Haufler) and the JSPS KEKENHI (JP20H01495, JP22K13388, JP24H00014) (Kato) is gratefully acknowledged. This study is conducted as a part of the Project ``Economic Policy Issues in the Global Economy'' undertaken at the Research Institute of Economy, Trade and Industry (RIETI). Kato thanks the University of Munich for their hospitality while part of this paper was written. We thank Tomohiro Ara, Real Arai, Ruby Doeleman, Antoine Ferey, Emanuel Hansen, Jim Hines, Niels Johannesen, Makoto Hasegawa, Jota Ishikawa, Mick Keen, Sang-Hyun Kim, Yoshimasa Komoriya, Jonas L{\"o}bbing, Mohammed Mardan, Jakob Miethe, Kenichi Nishikata, Hirofumi Okoshi, Pascalis Raimondos, Dirk Schindler, Karl Schulz, Georg Thunecke, and Alfons Weichenrieder for many helpful comments and discussions. Thanks also to seminar and conference participants at the IIPF in Utah State, MPI Berlin, Goettingen, Osaka, Kyoto, Shimane, QUT Brisbane, Ryukoku, Gakushuin, CESifo (Munich \& Venice Summer Institute), and RIETI for helpful comments.}}
\end{center}

\vspace{0.05cm}

\begin{center}
   {\large Andreas Haufler\renewcommand{\thefootnote}{\fnsymbol{footnote}}\footnote[2]{Corresponding author. Seminar for Economic Policy, University of Munich, Germany. {\it E-mail address:} andreas.haufler[at]econ.lmu.de.} \qquad Hayato Kato\renewcommand{\thefootnote}{\fnsymbol{footnote}}\footnote[3]{Graduate School of Economics, the University of Osaka, Japan. {\it E-mail address:} hayato.kato[at]econ.osaka-u.ac.jp.}}
\end{center}

\vspace{0.05cm}







\begin{abstract}
The Global Minimum Tax (GMT) is applied only to firms above a certain size threshold, permitting countries to set differential tax rates for small and large firms. We analyse tax competition among multiple tax havens and a non-haven country for heterogeneous multinationals to evaluate the effects of this partial coverage of GMT. Upon the introduction of a low but binding GMT rate, the havens commit to the single uniform GMT rate for all multinationals. However, gradual increases in the GMT rate induce the havens, and subsequently the non-haven, to adopt discriminatory, lower tax rates for small multinationals. Our calibration exercise shows that introducing a GMT rate close to 15\% results in a regime where only the havens adopt split tax rates. Welfare and tax revenues fall in the havens but rise in the non-haven, yielding a positive net gain worldwide.

\vspace{0.1cm}


\vspace{0.1cm}

\end{abstract}


\

\section{Introduction}
\renewcommand{\thefootnote}{\arabic{footnote}}
\setcounter{footnote}{0}

Profit shifting by multinational enterprises (MNEs) remains a central challenge for international corporate taxation. Aggregate estimates suggest that more than one-third of foreign-earned corporate profits are shifted to tax havens (\citealp{Torslovetal2023}).\footnote{Combining micro- and macro-level data on portfolio investment, \cite{Coppolaetal2021} highlight the role of tax havens as conduits through which firms access international capital markets. See \cite{Camareroetal2025} for an application of their dataset in the context of the EU.} The important role of tax havens for profit shifting is confirmed in other studies using microdata of multinational affiliates in France (\citealp{Daviesetal2018}), the U.K. (\citealp{Bilicka2019}), and the U.S. (\citealp{Guvenenetal2022}). In response to the large revenue losses caused by profit shifting, the OECD has launched an action plan to fight base erosion in OECD countries, and in particular the profit shifting to tax havens (\citealp{OECD2013}). A core development in this endeavor is the introduction of a global minimum tax (GMT), also known as {\it Pillar Two}.\footnote{See \cite{FuestNeumeier2023}; \cite{Hebousetal2024} for the institutional background. Another outcome of the BEPS initiative is {\it Pillar One}, which reallocates taxing rights from the country of a MNE's residence to the market jurisdiction (\citealp{Mukunokietal2025}; \citealp{OgawaTsuchiya2025}; \citealp{Richter2025}). Pillar One aims to address the tax challenges posed by the digital economy (see \citealp{Lassmannetal2025} for how multinational digital platforms respond to corporate taxes).} In 2021, more than 130 countries including many tax havens agreed on a 15\% GMT rate; by August 2025, 65 countries had drafted or enacted legislation implementing the GMT.\footnote{See Tax Foundation, `The Latest on the Global Tax Agreement': \url{https://taxfoundation.org/blog/global-tax-agreement/} (accessed January 29, 2026).
The United States had already implemented, in its 2017 Tax Cuts and Jobs Act, a tax on Global Intangible Low-Tax Income (GILTI) that works in many respects like a GMT  (\citealp{Chodorowetal2024}).}

An important limitation of the GMT is that it applies only to `large' multinationals, defined as groups with at least 750 million EUR in annual revenues in two of the last four years.\footnote{Since 2016, such groups have been required to file {\it country-by-country reports} (CbCR) on their activities, profits, and taxes in each country of operation (\citealp[Section 2]{Doelemanetal2024}). The revenue threshold reflects the balance between limiting profit shifting and minimising compliance costs \citep[p.14]{OECD2022}.}
Fig.~1 shows that, according to the Orbis database of Bureau van Dijk, about 30\% of MNEs, accounting for roughly 90\% of total profits, exceeded this threshold in 2018--2021 and were thus covered by the GMT.\footnote{This figure aligns with the OECD's estimate of 90\% profit coverage (\citealp[\# 505, p.~233]{OECD2020impact}).}
Nevertheless, over 300 billion EUR of MNEs' profits remain outside its scope. Moreover, Orbis tends to oversample large firms \citep{Bajgaretal2020}, and hence the 90\% coverage rate should be seen as an upper bound.

One implication of this incomplete coverage is that low-tax countries might respond to the introduction of the GMT by using a split corporate tax system, where a lower tax rate applies to firms below the GMT threshold. In fact, such a split is inherent in the regulation: tax havens can implement the GMT by raising their (low) base tax rates only for MNEs above the threshold through a specific top-up tax.\footnote{This is known as the Qualified Domestic Minimum Top-Up Tax (QDMTT). See \cite{Devereux2023} for a discussion. \label{footQDMTT} } Tax havens are fully aware of this option. Ireland, for example, has decided to keep its general tax rate at 12.5\%, but top up the tax rate to 15\% for affiliates of foreign MNEs above the threshold (\citealp{GovIreland2023}). Several other major low-tax countries, including Bahrain, Bulgaria, Hungary, Switzerland and the UAE have adopted similar split tax systems by joining the GMT agreement while maintaining their base statutory corporate tax rates below 15\% \citep{PwC2025}.\footnote{The latest corporate tax rates around the world can be found in `Tax Foundation, `Corporate Tax Rates Around the World, 2025': \url{https://taxfoundation.org/data/all/global/corporate-tax-rates-by-country-2025/} (accessed January 29, 2026). Note that not all low-tax countries have adopted a split tax system: Cyprus and Gibraltar, for example, raised their base rates from 12.5\% to 15\% for both large and small MNEs, effective in 2026 and 2024, respectively. See Shanda Consult, `The Cyprus Tax Reform 2026: A Strategic Evolution for Business and Compliance': \url{https://shandaconsult.com/cyprus-tax-reform-2026} and PwC, `Gibraltar---Taxes on corporate income': \url{https://taxsummaries.pwc.com/gibraltar/corporate/taxes-on-corporate-income} (accessed January 29, 2026).}

\begin{figure}
\begin{center}
\vspace{.2cm}

\includegraphics[scale=0.45]{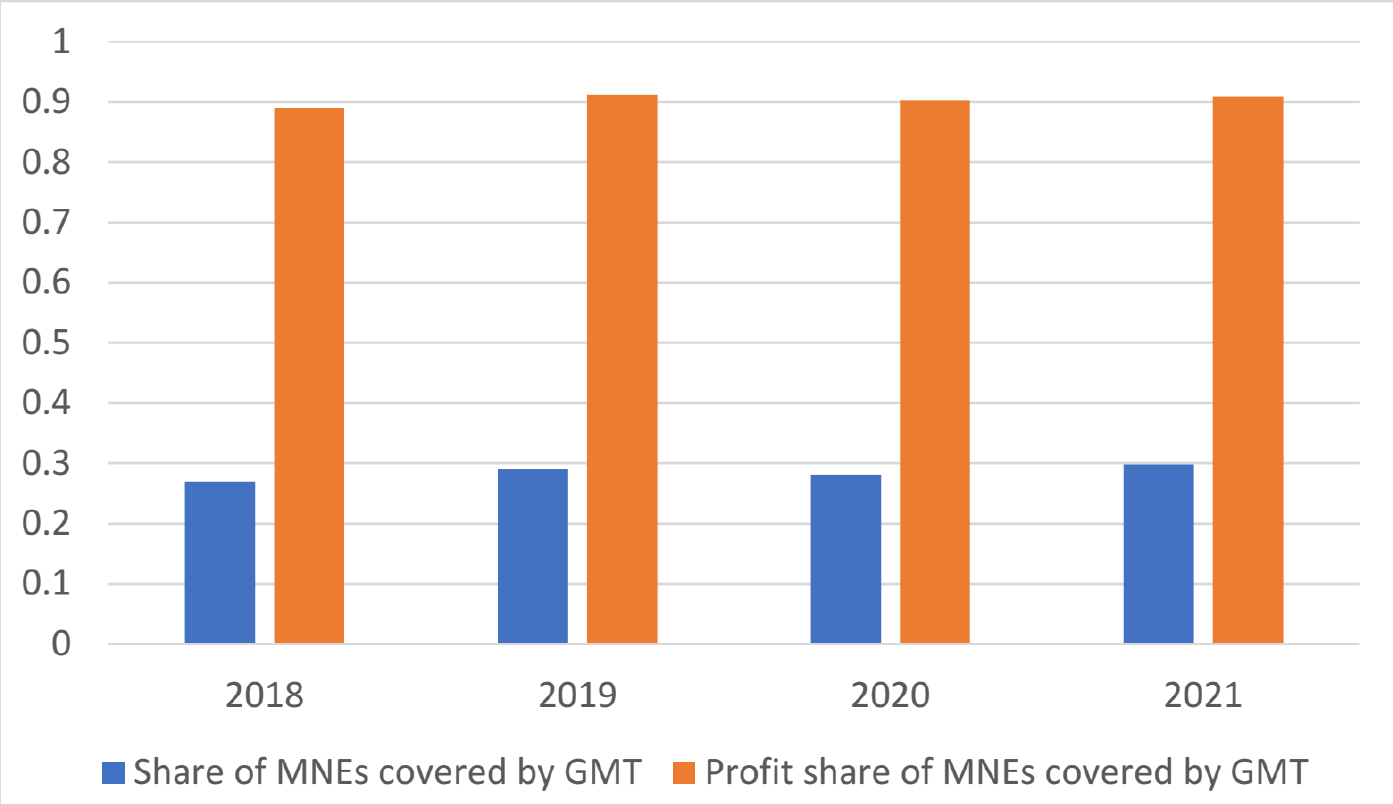} \\ 
\end{center}
\noindent
{\bf Fig. 1.} GMT coverage. \\
{\small \noindent {\em Source}: Orbis database, own calculations. \\ \noindent {\em Notes}: The left bar in each year  shows the share of MNEs covered by the GMT out of all MNEs.
The right bar in each year shows the share of pre-tax profits earned by MNEs covered by GMT out of pre-tax profits by all MNEs.
The MNEs covered by GMT are those whose annual revenues exceed $750$ million EUR in at least two of the last four years out of all MNEs.
See \ref{app: fig.1} for details.}

\vspace{-.3cm}
\end{figure}

These responses show that the effects of the GMT cannot be analysed assuming a single, uniform corporate tax rate across all countries. 
Against this background, this paper asks when tax competition leads countries to differentiate tax rates for large and small MNEs under the GMT, whether a 15\% GMT rate induces such a split, and how large are the resulting revenue and welfare gains. 

To answer these questions, we build a simple model where heterogeneous MNEs headquartered in a non-haven country shift profits to tax-haven countries. Countries maximise welfare---a weighted sum of tax revenues and, for the non-haven, the post-tax income of MNEs' owners---and compete for the MNEs' profits. In the first stage, both the non-haven and the havens decide whether to commit to a uniform tax rate for {\em all} MNEs. In the second stage, countries set tax rates, either uniform or split, with large MNEs taxed at least at the GMT rate. The final stage involves MNEs’ profit-shifting decisions.

Our analysis shows that introducing the GMT with partial coverage generates a sequence of non-cooperative tax equilibrium regimes between the non-haven and the havens. When the GMT rate is binding but low, the havens commit to it for all MNEs to raise the non-haven’s uniform tax rate. As the GMT rate rises, the havens start splitting their rates, applying the GMT to large MNEs and a lower rate to small MNEs. For sufficiently high GMT rates, the non-haven also splits its rate, as limited competition for large MNEs' profits shifts the focus towards small MNEs.

We then calibrate our model using estimates of profit shifting and examine the quantitative effects of GMT reforms. For our basic model, a 15\% GMT rate leads both the non-haven and the havens to choose uniform rates for all firms, whereas a 16\% GMT rate induces the havens to split their tax rates. In an extended model where tax havens choose their tax rates in a decentralized way, a 15\% GMT rate is sufficient to induce a split tax system for the havens in equilibrium. This splitting result is consistent with the observed responses in several tax havens, including Ireland and Switzerland, as discussed earlier. Our quantitative results also show that introducing a 15\% GMT rate benefits the non-haven while harming the havens, generating net global revenue gains of 149--156 billion USD (and welfare gains of 92--96 billion USD). These revenue gains are comparable to existing estimates in the literature \citep{Barakeetal2022, Huggeretal2024}. 


Our analysis contributes to the recent theoretical studies on tax competition under a GMT. They ask how introducing a GMT affects national and global revenues and welfare in the presence of country heterogeneity \citep{Johannesen2022, HebousKeen2023}; real economic activities \citep{JanebaSchjelderup2023, SchjelderupStaehler2024}; and endogenous tax-enforcement levels \citep{HindriksNishimura2025}. All these models assume the GMT applies uniformly, ignoring the firm-size threshold below which firms are exempt from the GMT. Our model highlights this threshold, allowing countries to tax firms above and below it differently.\footnote{\label{fn: differential}This links our model to the literature on discriminatory tax competition, where countries can tax firms differently by their mobility (\citealp{JanebaPeters1999}; \citealp{Keen2001}; \citealp{Peraltaetal2006}; \citealp{GaigneWooton2011}; \citealp{HauflerRunkel2012}).}

Our simple quantitative exercise yields results that are comparable to more detailed estimates of the revenue effects of a GMT \citep{Barakeetal2022,Huggeretal2024}. It also complements recent evaluations of corporate taxation using trade and multinational models \citep{Wang2020, Ferrarietal2023, Shen2024, Dyrdaetal2024, Santacreu2026}, where taxes are assumed to be exogenous and/or uniform across firms.
By contrast, we abstract from firms' real activities and focus on the strategic role of size-dependent taxes. One of the few papers offering empirical evidence on minimum tax regimes is \cite{BuettnerPoehnlein2024}, who study the effects of introducing a minimum local business tax rate in Germany. Finally, two recent empirical papers \citep{Bachasetal2025,Cliffordetal2025} examine which firms fall within the scope of the GMT using firm/affiliate level data for 16 countries and Germany, respectively. Their findings that GMT revenue gains depend strongly on its coverage emphasize the relevance of our analysis.

We proceed as follows. In Section~\ref{sec: model} we set up our model of tax competition. Section~\ref{sec: regimes} characterises the sub-game perfect Nash equilibria that result in different tax regimes. Section~\ref{sec: welfare} turns to the welfare effects of introducing a GMT, and of increasing its tax rate and coverage. Section~\ref{sec: quantification} calibrates our model to real-world data and quantifies its effects. Section~\ref{sec: conclusion} concludes.

\section{The model}\label{sec: model}

\subsection{Setup}

We consider tax competition  between a non-haven country, indexed by $n$, and a finite set of $H$ symmetric tax-haven countries, indexed by $h$. 
There is a continuum of heterogeneous MNEs which are headquartered in the non-haven and have affiliates in all havens.\footnote{To focus on the aggregate level of profit shifting relevant for countries’ corporate tax revenues, we abstract from the endogenous location choice of MNEs. On the evolution of US-owned MNEs over time and space, see \cite{Guvenenetal2022} for profit shifting and \cite{Garettoetal2024} for real economic activities.} Headquarters earn (`true') exogenous profits $\pi \in [\pu, \infty)$ with distribution $F(\pi)$, arising from real activity in the non-haven.
MNEs can shift some of these profits to the havens to maximise their post-tax profits. In the following, we assume (i) true profits $\pi$ independent of taxes and (ii) the collective decision-making by the havens for analytical simplicity. These assumptions are relaxed in the quantitative analysis in Section~\ref{subsec: extensions}, where we explore the robustness of our results.

We assume the non-haven and all havens have signed the GMT agreement and must set tax rates no lower than the GMT rate $t_M$.\footnote{\cite{Devereux2023} analyses how the GMT’s institutional design, especially the Under-Taxed Payments Rule (UTPR), incentivizes non-haven and haven countries to join once a critical mass of large non-havens participates.} The GMT applies only to MNEs with profits $\pi \ge \pi_M$ (`large MNEs'), allowing countries to set lower rates for firms with $\pi < \pi_M$ (`small MNEs').

The non-haven maximises a welfare function that consists of a weighted sum of tax revenues and the post-tax profits of MNEs, which are all owned by residents of the non-haven. Since havens do not have any ownership in the firms' profits, the welfare of tax havens corresponds to their tax revenues.

We analyse a three-stage game. In the first stage, the non-haven and havens simultaneously decide whether to commit to a single tax rate for all MNEs that is at or above the GMT rate, or to allow differentiated rates.\footnote{The ability to commit, a realistic feature of international tax relations, is standard both in the studies cited in footnote~\ref{fn: differential} and in related work on endogenous timing in games \citep{Hucketal2002, KempfRotaGraziosi2010}. The choice between a uniform and a split tax system is more fundamental, and less frequently revised, than the choice of tax rates.} The havens decide collectively on whether to unify or split their rates. In the second stage, countries set their tax rates non-cooperatively. Those that committed choose a single rate; others set different rates for large and small MNEs. In the final stage, MNEs shift profits given these tax rates. The game is solved by backward induction, yielding a sub-game perfect Nash equilibrium. We also refer to this as the tax-competition equilibrium.

\subsection{Profit shifting by multinationals}

Each MNE shifts a share $\theta_h$ of its pre-tax profits $\pi$ from its non-haven headquarters to each tax haven affiliate $h$, and incurs transaction costs $C_h(\cdot)$ for doing so. These costs can be interpreted as concealment costs or as expected fines, if detected. We assume the costs are quadratic in the level of profit shifting but decrease with the size of a firm: $C_h(\theta_h \pi)=\delta(\theta_h \pi)^2/(2\pi)=\delta \theta_h^2 \pi/2$, resulting in total shifting costs of $\sum_{h=1}^H C_h(\theta_h \pi)$.\footnote{This specification is standard in the literature (e.g., \citealp{HinesRice1994}; \citealp{HuizingaLaeven2008}; \citealp{Suarez2018}; \citealp{Bilickaetal2024}).} The parameter~$\delta$ captures the difficulty of profit shifting.

Given the tax rates $t_n$ and $t_h$ respectively in the non-haven and each of the havens, an MNE maximises its global post-tax profits net of profit-shifting costs, denoted by $\widetilde \pi$, by choosing the share of profit shifting $\theta_h$ to each haven:\footnote{The GMT tax base equals that of national corporate taxes. We thus ignore additional deductions (`substance-based income exclusion'), an issue examined by \cite{SchjelderupStaehler2024}.}
\begin{equation}
\max_{\{ \theta_h \}_{h=1}^H } \ \widetilde{\pi}\left( \{ \theta_h \}_{h=1}^H; \pi\right) = (1-t_n)\left( 1 -\sum_{h=1}^H \theta_h \right)\pi + \sum_{h=1}^H (1-t_h)\theta_h \pi - \sum_{h=1}^H \frac{\delta (\theta_h \pi)^2}{2 \pi}. \label{MNEmax}
\end{equation}
The optimal choice is
\begin{equation}
\theta_h = \frac{t_n - t_h}{\delta}, \label{MNEFOC}
\end{equation}
which is independent of $\pi$ and therefore holds for all MNEs.

The optimal $\theta_h$ in \eqref{MNEFOC} implies that MNEs of different sizes share the same tax base elasticity. Substituting \eqref{MNEFOC} into \eqref{MNEmax} and using the symmetry of havens, the tax base of an MNE with $\pi$ in the non-haven becomes $TB_n = [1 - H(t_n - t_h)/\delta]\pi$, yielding $\varepsilon_n = - \frac{dTB_n/TB_n}{dt_n/t_n} = \frac{t_n}{\delta - H(t_n - t_h)}$, which is independent of $\pi$. Equal tax base elasticities thus remove countries' incentives to tax-discriminate between MNEs in the {\em absence} of the GMT. This allows us to focus on the splitting responses to a GMT that is applied only to large firms.

The empirical evidence on the relationship between the size of firms and their responsiveness to tax is not conclusive. Media coverage of tax avoidance by very large multinationals gives an impression that bigger firms have a higher tax-base elasticity. However, there are several studies which find the opposite result that the tax-base elasticity is indeed higher for small firms (not necessarily MNEs).\footnote{Making use of bunching in the distribution of taxable income of firms in the UK, \cite{Devereuxetal2014} find that small-sized firms change their taxable income more significantly in response to statutory tax rate changes, as compared to medium-sized firms. Applying a similar empirical strategy to US firms, \cite{Colesetal2022} find that the elasticity of taxable income with respect to effective tax rates is monotonically decreasing in firm size. See also \cite{Auliffeetal2023} and the references therein for the heterogeneous tax elasticities of tangible asset investments among firms of different sizes.}  
A detailed analysis of French MNEs using Country-by-Country Reports (CbCR) empirically supports our result in eq.~\eqref{MNEFOC} that firms of different size shift roughly the same {\em share} of their profits (\citealp{Aliprandietal2025}, Figure 5). This is compatible with the finding that large MNEs are responsible for most of the {\em  total} amount of profit shifting (\citealp{Cliffordetal2025}), given their large share of total profits. In summary, assuming the tax base elasticity to be independent of firm size seems to be a useful and empirically plausible baseline.


\subsection{Governments' tax setting choices}

Before examining the effects of the GMT, we first solve the {\em unconstrained} optimisation problem in the absence of the GMT. The non-haven country~$n$ maximises a weighted sum of private income and tax revenues, where private income are the profits (net of taxes and profit-shifting costs) of all MNEs, whose owners reside exclusively in the non-haven. The non-haven's tax revenues are weighted with a factor $\lambda >1$, which represents the marginal value of public funds (MVPF) (\citealp{HebousKeen2023}). For $\lambda \rightarrow \infty$, we obtain the special case of a Leviathan government that is solely interested in maximising its tax revenues (e.g., \citealp{JanebaSchjelderup2023}; \citealp{HindriksNishimura2025}).

The government of the non-haven~$n$ therefore maximises {\small
\begin{align} \label{Gn}
G_n &=\underbrace{\int_{\underline{\pi}}^{\infty}  \widetilde{\pi}\left( \{ \theta_h \}_{h=1}^H; \pi\right) dF}_{\text{Private benefit}} + \lambda \underbrace{\int_{\underline{\pi}}^{\infty} t_n\left( 1 -\sum_{h=1}^H \theta_h \right)\pi dF}_{\text{Tax revenues}} \nonumber \\
\
&= (1-t_n)\left( 1 - \frac{Ht_n - \sum_{h=1}^H t_h}{\delta}\right)\Pi + \sum_{h=1}^H (1-t_h) \left( \frac{t_n-t_h}{\delta}\right) \Pi - \sum_{h=1}^H \frac{\delta}{2}\left( \frac{t_n-t_h}{\delta} \right)^2 \Pi \nonumber \\
&\qquad \qquad + \lambda t_n \left( 1 - \frac{Ht_n - \sum_{h=1}^H t_h}{\delta}\right)\Pi,
\end{align} } where $\Pi \equiv \int_{\pu}^{\infty}\pi dF$ are the total true profits of MNEs.

Each of the havens $h$ only cares about its tax revenues (multiplied by the MVPF $\lambda$), as the affiliate located in $h$ is owned by its headquarters in the non-haven. This gives
\begin{equation}
  G_h = \lambda \int_{\underline{\pi}}^{\infty} t_h \theta_h \pi dF = \lambda t_h \left( \frac{t_n-t_h}{\delta} \right)\Pi. \label{Wh}
\end{equation}
To determine world welfare, we take a utilitarian approach and define $G_W \equiv G_n + \sum_{h=1}^H G_h$, using~\eqref{Gn} and~\eqref{Wh}.

Solving the first-order conditions of the two sets of countries yields best responses:\footnote{The second-order conditions of both countries' optimal tax problems are trivially satisfied.}
\begin{equation}
    t_n = \frac{(\lambda-1)\left( \sum_{h=1}^H t_h +\delta\right)}{H(2\lambda-1)}, \qquad t_h = \frac{t_n}{2}. \label{response}
\end{equation}
Solving for the optimal tax rates in the unconstrained benchmark, which we call `Regime 0,' gives
\begin{equation}
    t_n^0 = \frac{2\delta(\lambda-1)}{H(3\lambda-1)}, \qquad t_h^0 = \frac{\delta(\lambda-1)}{H(3\lambda-1)}. \label{UTAX}
\end{equation}
We hereafter assume $\delta \in (0, \, 3H/2)$ to ensure $t_n^0 \in (0, 1)$.

Two comments on \eqref{UTAX} are in order.
First, tax rates in all countries are rising in the profit-shifting costs $\delta$ and falling in the number of tax havens $H$. Second, they are rising in the MVPF, $\lambda$. A higher value of $\lambda$ raises the optimal tax rate in the non-haven by increasing the value of tax revenues vis-\`{a}-vis the post-tax profits of MNEs. The higher tax rate in the non-haven in turn leads each of the havens to raise its tax rate as well.\footnote{\label{foot: johannesen}Note the difference between this result and the analysis in \cite{Johannesen2022}, where the tax rate in each of the tax havens is zero in the absence of the GMT. \cite{Johannesen2022} assumes that shifting profits from a high-tax to a low-tax country incurs zero cost so that MNEs shift their profits only to the lowest-tax country. By contrast, we assume that profit shifting from the high-tax country to a given low-tax country has convex costs; hence MNEs diversify their allocation of shifted profits across all havens.}

The share of shifted profits in Regime~0 is given by
\begin{equation}
    \sum_{h=1}^H \theta_h^0 = H \cdot \left( \frac{t_n^0-t_h^0}{\delta}\right) = \frac{H}{\delta} \cdot \frac{\delta(\lambda-1)}{H(3\lambda-1)} = \frac{\lambda-1}{3\lambda-1} < \lim_{\lambda \to \infty} \sum_{h=1}^H \theta_h^0 = \frac{1}{3}.   \label{shift}
\end{equation}
Hence, when the non-haven maximises tax revenues only ($\lambda \to \infty$), the international tax differential between the non-haven and the havens is maximal, resulting in one-third of all profits of MNEs being shifted to the havens in equilibrium. For lower values of $\lambda$, the equilibrium tax differential is reduced, and the share of shifted profits is accordingly lower.









\section{Equilibrium tax regimes} \label{sec: regimes}

With a GMT in place, in the first stage of the game each country decides whether to commit to a single tax rate---equal to or above the GMT rate. Non-committing countries split rates in the second stage, setting different rates for large and small MNEs. Large MNEs with profits $\pi \in [\pi_M, \infty)$ are subject to the GMT, while small MNEs with $\pi \in [\underline{\pi}, \pi_M)$ are not. Let $\phi$ denote the exogenous share of profits earned by large MNEs, i.e., the {\em coverage rate} of the GMT:
\begin{equation} \label{phi}
\phi \equiv \dfrac{\int_{\pi_M}^{\infty} \pi dF}{\int_{\pu}^{\infty} \pi dF}
\equiv \dfrac{\int_{\pi_M}^{\infty} \pi dF}{\po}.
\end{equation}

\begin{table}[!ht]
\begin{center}
\begin{flushleft}
{\bf Table 1} \\
Possible regimes of the non-cooperative tax game.
\end{flushleft}
\vspace{0.1cm}
    \setlength{\extrarowheight}{2pt}
    \begin{tabular}{|l||c|c|c|}
       \hline {\bf Non-haven}  & \multicolumn{3}{c|}{\bf Haven}\\
       \multicolumn{1}{|c||}{} & \multicolumn{1}{c|}{single non-GMT rate}  & \multicolumn{1}{c|}{single GMT rate} &  \multicolumn{1}{c|}{split tax rate} \\  \hline \hline
       single non-GMT rate & Regime 0 & Regime 1  & Regime 2 \\ \hline
       single GMT rate & ---  & --- &  Regime 3 \\   \hline
       split tax rate  & --- & ---  & Regime 4 \\ \hline
    \end{tabular}
\end{center}
\end{table}

Table~1 shows possible combinations of tax policies in the non-haven and the havens. Of the nine potential tax regimes, four regimes cannot occur in equilibrium. In these regimes, if the non-haven adopts a single GMT rate, the havens set a uniform tax rate at the same level or higher and therefore earn no positive tax revenues. Clearly, this cannot be an equilibrium because each haven can secure strictly positive tax revenues by splitting its tax rate and underbidding the tax rate of the non-haven for small MNEs. Alternatively, if the havens set a single non-GMT or a single GMT rate, the non-haven never wants to split its tax rate, due to the constant tax base elasticity. Hence only five possible tax regimes remain.


Regime 0 is the unconstrained tax-competition equilibrium, where the GMT rate $t_M$ is non-binding in all countries, as discussed earlier. As $t_M$ rises, the equilibrium moves through four additional regimes, which are characterised as follows.

\

\begin{Proposition}

\noindent {\bf (Equilibrium regimes with binding GMT)} \quad Consider a GMT rate $t_M \ge t_M^0 \equiv t_h^0$, a GMT coverage rate $\phi \in (0, 1)$, a marginal value of public funds $\lambda >1$, and a cost parameter of profit shifting $\delta \in (0, 3H/2)$. As $t_M$ is continuously increased, the tax-competition equilibrium is characterised by the following four regimes, with all symmetric havens choosing the same tax schedule.


\begin{itemize}
\item[(i)] Regime 1: $t_M \in \left[ t_M^0, t_M^1 \equiv \frac{\delta(\lambda-1)(2\lambda-1)}{H[\lambda(3\lambda-1) -\phi(\lambda-1)^2]} \right]$. \quad The non-haven chooses a single non-GMT rate, and the (representative) haven chooses a single GMT rate:
\begin{equation*}
t_n^1 = \frac{(\lambda-1)(Ht_M+\delta)}{H(2\lambda-1)} \ \ \text{for} \ \pi \in [\pu, \infty), \qquad t_h^1 =
t_M \ \ \text{for} \ \pi \in [\pu, \infty).
\end{equation*}
\item[(ii)] Regime 2: $t_M \in \left( t_M^1, t_M^2 \equiv \frac{2\delta(\lambda-1)}{H[3\lambda-1 -\phi(\lambda-1)]} \right]$. \quad The non-haven sets a single non-GMT rate, and the haven splits its tax rate and chooses the GMT rate for large MNEs, but a lower rate than the GMT for small MNEs: {\small
\begin{equation*}
t_n^2 = \frac{2(\lambda-1)(\phi H t_M +\delta)}{H[\lambda(3+\phi) -(1+\phi)]} \ \ \text{for} \ \pi \in [\pi_M, \infty), \qquad t_h^2 = \begin{cases}
\frac{(\lambda-1)(\phi H t_M +\delta)}{H[\lambda(3+\phi) -(1+\phi)]} \ \ &\text{for} \ \pi \in [\pu, \pi_M) \\
t_M \ \ &\text{for} \ \pi \in [\pi_M, \infty)
\end{cases}.
\end{equation*} }
\item[(iii)] Regime 3: $t_M \in \left( t_M^2, t_M^3 \equiv \frac{2\delta(\lambda-1)(8\lambda-3)}{H(3\lambda-1)(4\lambda-1)} \right]$. \quad The non-haven chooses a single GMT rate, and the haven splits its tax rate and chooses a GMT rate for large MNEs, but a lower rate than the GMT for small MNEs:
\begin{align*} \label{t3}
\begin{split}
t_n^3 = t_M \ \ \text{for} \ \pi \in [\pu, \infty), \qquad
t_h^3 =\begin{cases} t_M/2 &\text{for} \ \pi \in [\pu, \pi_M) \\
t_M &\text{for} \ \pi \in [\pi_M, \infty)
\end{cases}.
\end{split}
\end{align*}
\item[(iv)] Regime 4: $t_M \in (t_M^3, 1]$. \quad Both the non-haven and the haven split their tax rates and choose the GMT rate for large MNEs, but a lower tax rate for small MNEs:
\begin{equation*}
\begin{split}
t_n^4 = \begin{cases} \frac{2\delta(\lambda-1)}{H(3\lambda-1)} &\text{for} \ \pi \in [\pu, \pi_M) \\
t_M &\text{for} \ \pi \in [\pi_M, \infty)
\end{cases}, \qquad
t_h^4 =\begin{cases} \frac{\delta(\lambda-1)}{H(3\lambda-1)} &\text{for} \ \pi \in [\pu, \pi_M) \\
t_M &\text{for} \ \pi \in [\pi_M, \infty)
\end{cases}.
\end{split}
\end{equation*}
\end{itemize}

\noindent Proof: See \ref{app: prop01}.
\end{Proposition}

\vspace{.3cm}

In the following we go through the different regimes, which are divided by the three threshold GMT rates $\{ t_M^i\}_{i=1,2,3}$. We call these the {\em regime-switching rates}.

In Regime~1, when the GMT rate $t_M$ exceeds the havens' unconstrained rate $t_h^0$, they must apply $t_M$ to large MNEs. The non-haven's unconstrained rate remains above $t_M$, so it continues to set a single rate above $t_M$ for all MNEs. The havens then choose whether to apply $t_M$ also to small MNEs or use a split schedule. By committing to a single GMT rate, they can induce the non-haven to raise its uniform rate. This argument is strongest when the havens decide collectively, as assumed in our main model.\footnote{\label{fn: decentralised}Qualitatively, the same results hold when the havens make their first-stage decisions in a decentralised manner, as shown in Section~\ref{subsec: extensions}. In this case, the parameter range of Regime~1 is smaller, since the commitment of a single haven induces the non-haven to raise its tax rate less than the simultaneous commitment of multiple havens.} The havens trade off gains from inducing a higher non-haven rate against revenue losses from not undercutting the non-haven's rate even more for small MNEs. In Regime~1, the first effect dominates, so they commit to the GMT rate for all MNEs, leading all countries to set higher uniform tax rates than in Regime~0.

In Regime~2, with $t_M$ above the regime-switching rate $t_M^1$, the non-haven remains unconstrained and sets a single rate above $t_M$ for all MNEs. For the havens, however, the trade-off described in Regime~1 above changes. The higher $t_M$ reduces gains from a uniform rate, as the tax differential and thus shifted profits shrink.\footnote{This is seen from Proposition~1(ii), where the tax differential $t_n^2 -t_h^2 = [\delta(\lambda-1)-\lambda Ht_M]/[H(2\lambda-1)]$ decreases with $t_M$.} At the same time, it increases losses from not competing with the non-haven for small MNEs. Consequently, the havens split their rates, applying $t_M$ to large MNEs and a lower rate to small MNEs.

In Regime~3, with $t_M$ above the regime-switching rate $t_M^2$, the non-haven can no longer set a rate above $t_M$ but is bound by it for large MNEs. By committing to $t_M$ for all MNEs, it induces the havens to raise their rates for small MNEs, analogous to the argument in Regime~1. In equilibrium, the non-haven applies $t_M$ to all firms, while the havens set $t_M/2$ for small MNEs.


Finally, in Regime 4, with $t_M$ above the regime-switching rate $t_M^2$, the international tax differential for small MNEs becomes very large, if the non-haven keeps its commitment to set $t_M$ for all MNEs.\footnote{This is seen from Proposition~1(iii), where the tax differential $t_n^3 -t_h^3 = t_M/2$ increases with $t_M$.}
To limit tax base losses from small MNEs, the non-haven abandons its commitment and lowers its tax rate for them. Again, this argument is analogous to the one for the havens to end their commitment in Regime~2. In Regime~4, both countries split their rates, applying unconstrained rates to small MNEs and the GMT rate to large MNEs.


\begin{figure}[!ht]

\vspace{.3cm}

\begin{center}
\includegraphics[scale=0.75]{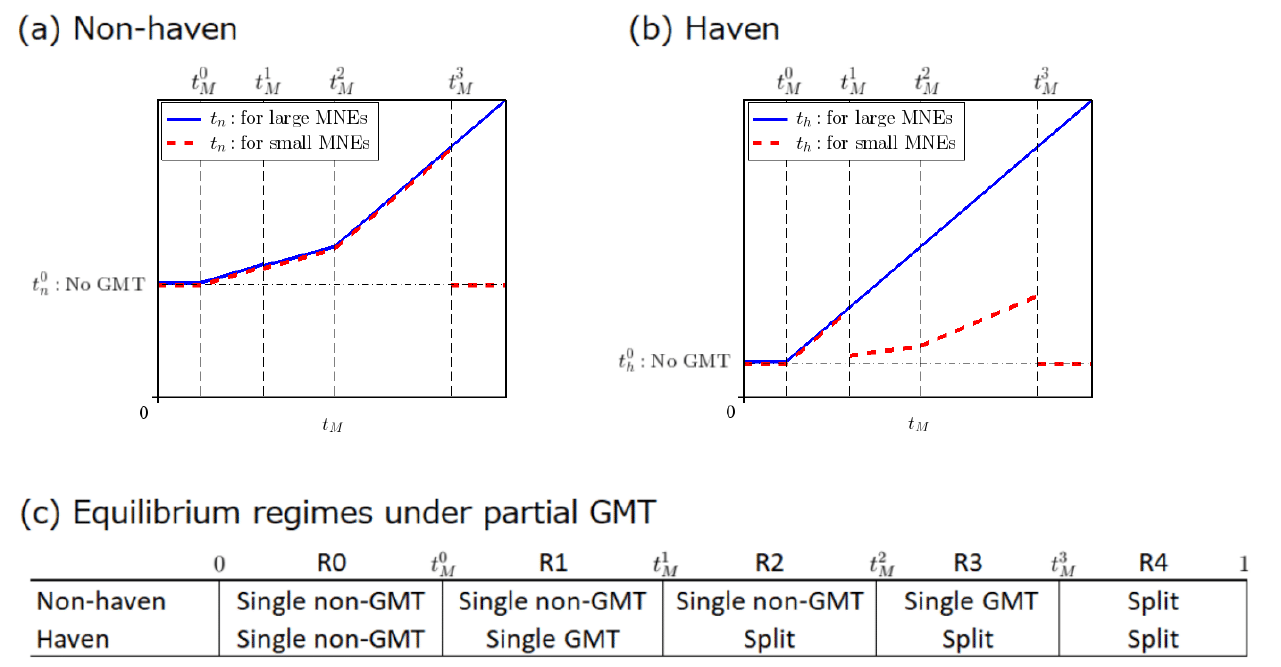} \\
\end{center} \vspace{-0.2cm}
\noindent {\bf Fig. 2.} Equilibrium tax rates for different GMT rates. \\
\noindent {\small {\it Notes}: The figure shows the equilibrium tax rates in the non-haven in panel (a) and the haven in (b) for different GMT rates $t_M$, and the summary of equilibrium regimes in (c). While the levels of tax rates and regime-switching rates vary with parameter values, the qualitative patterns remain unchanged.}
 \vspace{-.3cm}
\end{figure}

Fig.~2 shows the equilibrium tax rates in the different regime-specific equilibria for varying values of the GMT rate $t_M$. The non-haven's tax rates are given in panel (a), while the havens' tax rates are in panel (b). The equilibrium regimes are summarised in panel (c). The solid (dotted) curves are the tax rates for large MNEs (small MNEs) in the tax-competition equilibrium under the GMT; the dashed horizontal lines are the unconstrained equilibrium tax rate $t_i^0$ (in Regime 0). Tax rates in both countries generally rise as $t_M$ increases, but there are distinct patterns in the different regimes. At the switch from Regime~1 to 2, the havens begin to split their tax rates, and the rate for small MNEs falls discontinuously. At the switch from Regime~3 to 4, the non-haven splits its tax rate and discretely cut its tax rate for small MNEs to the level in the unconstrained Regime~0. The havens respond to that by also discontinuously reducing their tax rates on small MNEs to the Regime~0 level.

\section{Welfare effects of the GMT}\label{sec: welfare}

Having fully described the tax-competition equilibrium in each regime, we can now turn to the welfare effects of introducing the GMT.  As in Proposition 1, the qualitative nature of the propositions below holds independently of the marginal valuation of public funds, $\lambda$, and it includes revenue maximization ($\lambda \rightarrow \infty$) as a special case. Our discussion of the results will therefore focus on the implications that the GMT has on tax revenues.


We compare welfare levels in the tax-competition equilibrium to those in the unconstrained equilibrium, Regime~0. The results are summarized in:

\vspace{.3cm}

\begin{Proposition} {\bf (GMT introduction)} \quad Consider the tax-competition equilibrium with GMT, as summarized in Proposition~1,  and assume the GMT coverage rate is not too low so that $\phi \in (3/4, 1)$. Compared to the unconstrained equilibrium (Regime~0), the introduction of a binding GMT rate leads to:
\begin{itemize}
\item[(i)] a rise in welfare in the non-haven country for all Regimes 1 to 4.
\item[(ii)] a rise in welfare in the haven country for the first part of Regime 1 ($t_M < t_M^+ \equiv \frac{\delta(\lambda-1)(2\lambda-1)}{H\lambda(3\lambda-1)}$), but a fall in welfare in the rest of Regime~1 and in Regimes~2 to~4 ($t_M > t_M^+$).
\item[(iii)] a rise in world welfare for all Regimes 1 to 4.
\end{itemize}


\vspace{.15cm}

\noindent Proof:  See \ref{app: prop02}.
\end{Proposition}

Introducing a small GMT slightly higher than the havens' unconstrained rate $t_h^0$ leads to an equilibrium in Regime~1, with higher tax rates in both the non-haven and the haven countries. At the same time, profit shifting is reduced in comparison to the no-GMT benchmark due to a smaller tax differential. The sum of these effects must certainly benefit the non-haven. It also benefits the havens, as long as the gain from higher rates exceeds the loss of the havens' tax base, i.e., shifted profits.
However, once the GMT rate exceeds a certain threshold, $t_M^+$, which still lies in Regime~1, the reduced tax differential and thus the fall in shifted profits will dominate from the perspective of the haven countries. Therefore, for all levels of the GMT above $t_M^+$, tax havens lose from the introduction of the GMT while the non-haven gains.\footnote{This conflict of interest does not arise in \cite[Proposition~6]{Johannesen2022}, where havens gain from the introduction of a GMT at any level, as long as profit shifting is not completely eliminated. In his setting where MNEs shift profits only to the lowest-tax country, all the havens end up setting their tax rate to zero. Therefore, revenues in the unconstrained tax competition equilibrium are zero for each haven, and any positive GMT rate generates a revenue increase for the representative haven. See also footnote~\ref{foot: johannesen}.}

In the remaining Regimes 2 to 4, the tax differential for large MNEs is always smaller than the one in Regime~0.  As long as large MNEs covered by the GMT account for a sufficiently high share of aggregate profits, $\phi \in (3/4, 1)$, the havens lose from the introduction of $t_M$ higher than $t_M^+$ due to the reduced profit shifting of large MNEs. Hence, the effects of introducing the GMT on tax revenues and welfare in the non-haven and the haven countries remain opposed to each other throughout Regimes~2 to~4.\footnote{If small MNEs not covered by the GMT account for a sufficiently large share of aggregate profits so that $\phi \in (0, 3/4)$, then the  introduction of the GMT rate would instead benefit the havens in Regime~3. See \ref{app: prop02} for details.} Finally, the gains to the non-haven exceed the losses (if any) to the havens, so that the GMT introduction raises tax revenues and welfare worldwide in all regimes. These aggregate welfare gains result from the higher tax rates in both countries and the reduced levels of profit shifting. Fig.~C.1 in \ref{app: figures} illustrates the welfare effects summarised in Proposition~2. 

After the GMT has been introduced, two further reform options exist: (i) a marginal increase in the GMT rate, $t_M$, and (ii) a marginal increase in the GMT coverage rate, $\phi$. An increase in $t_M$ generally increases welfare in the non-haven and the world, except at the regime-switching rates where either of the non-haven or the havens introduce a split tax system. For the non-haven country, the welfare effects of  marginal increases in the GMT rate are non-monotonous. A higher $\phi$ has non-negative welfare effects for the non-haven and the world in all regimes, and it has non-positive welfare  effects for the havens. Intuitively, a rise in $\phi$ limits tax competition for small MNEs, and therefore amplifies the welfare effects arising from the GMT. We fully study these reforms in \ref{app: prop03} and \ref{app: prop04} respectively.

\section{Quantitative implications}\label{sec: quantification}

\subsection{Basic model}\label{subsec: basic} 
We calibrate the unconstrained model with no GMT (Regime~0) to match international profit-shifting data, as summarised in Table~2. The GMT coverage rate is set to $\phi=0.9$, following \cite{OECD2020impact} and our calculations in Table~1.  
The number of tax havens is set to $H=40$, as in \citet{Torslovetal2023}. Combining their data with Country-by-Country Reports (CbCR) yields total MNE pre-tax profits of $\Pi=4,623$ billion USD. The marginal value of public funds (MVPF), $\lambda$, and the cost parameter of profit shifting, $\delta$, are jointly calibrated to minimise the squared distance between the two model-predicted moments--- the non-haven tax rate $t_n^0$ and the shifted profit share $\sum_h \theta_h^0$---and their empirical counterparts.
For non-targeted moments, our model slightly underestimates the havens' tax rate, $t_h^0$, and the non-haven's revenue losses, $t_n^0 \sum_h \theta_h^0 \Pi$, but overall fits the data reasonably well.


\begin{table}[!ht]
\begin{flushleft} {\bf Table~2} \\
Calibration of the model.
\end{flushleft} \begin{center} {\small
\renewcommand{\arraystretch}{1.3}
\begin{tabular}{lll}
   \hline
   {\bf Parameter}  &  & Source \\ \hline \hline
   $\phi = 0.9$ & Profit share of MNEs covered by GMT & \cite{OECD2020impact}, Orbis \\
   $H = 40$ & Number of tax havens & \cite{Torslovetal2023} \\
   $\Pi = 4,623${\small bUSD} & Pre-tax profits of MNEs & \cite{Torslovetal2023}, CbCR \\
    $\lambda=2.1$ & Marginal value of public funds & Calibrated \\
   $\delta=17.8$ & Cost of profit shifting & Calibrated \\
   \hline
  \end{tabular} } \\ 
\end{center}
\begin{center}  {\small
\renewcommand{\arraystretch}{1.3}
\begin{tabular}{llll}
   \hline
   {\bf Targeted moment} & Model & Data & Source \\ \hline \hline
  Corp. tax rate of non-haven: {\footnotesize $t_n^0$} & $18.6\%$ & $18.6\%$ &  {\footnotesize \cite{Torslovetal2023}}  \\
  Share of shifted profits: {\footnotesize $\sum_h \theta_h^0$} & $20.9\%$ & $20.9\%$ &  {\footnotesize \cite{Torslovetal2023}}, CbCR \\
   \hline
  \end{tabular} } \\ 
  \end{center}
\begin{center}  {\small
\renewcommand{\arraystretch}{1.3}
\begin{tabular}{llll}
   \hline
   {\bf Non-targeted moment}  & Model & Data & Source \\ \hline \hline
   Corp. tax rate of haven: {\footnotesize $t_h^0$} & $9.3\%$ & $13.7\%$ & {\footnotesize \cite{Torslovetal2023}} \\
   Non-haven's revenue loss: $t_n^0 \sum_h \theta_h^0 \po$ & $180$bUSD & $247$bUSD & {\footnotesize \cite{Torslovetal2023}} \\
   \hline
  \end{tabular} } 
  \end{center}
  {\small {\it Notes}: Data are from 2019. The definitions of non-haven and haven countries follow \cite{Torslovetal2023} and the data is from the authors' website: \url{https://missingprofits.world} (corresponding figures in the spreadsheets of `Table 1' and `Table U1' in the excel file `1975-2019 updated estimates: Tables'). The non-haven countries are $30$ OECD countries, $7$ major developing countries and the rest of the world. There are $40$ haven countries across the world. The corporate tax rates for the representative the non-haven and haven countries are calculated as the GDP-weighted averages of their respective effective tax rates. \cite{Torslovetal2023} report the pre-tax profits of foreign-owned MNEs only, which do not include those of domestically-owned MNEs (see their Section 3.1.2 on p.7). Using the CbCR Statistics of the EU Tax Observatory (\url{https://www.taxobservatory.eu/repository/the-cbcr-explorer/}), we compute the pre-tax profit ratio of domestically-owned MNEs to foreign-owned MNEs. With this ratio ($0.785$) and the pre-tax profits of foreign-owned MNEs ($2,590$ billion USD), we obtain the total pre-tax profits of foreign- and domestically-owned MNEs as $(1+0.785)\times 2,590=4,623$ billion USD.}
\end{table}


Using this calibrated version of our model, we examine the quantitative effects of GMT reforms, as summarised in Table~3. The key result is that a 15\% GMT rate with partial coverage ($\phi=0.9$) yields an equilibrium in Regime~1 (Table~3(a)), whereas a 16\% GMT rate leads to an equilibrium in Regime~2 (Table~3(b)). At $t_M=0.16$, the havens split their tax rates and set a lower tax rate ($10.3\%$) for small MNEs as compared to large MNEs (16.0\%) (Proposition~1(ii)). Indeed, the threshold GMT rate is $t_M^1 = 0.156$, above which Regime~2 prevails. As we will see below, the threshold level for the GMT to induce an equilibrium in Regime~2 is further reduced, if the havens set their tax rates in a decentralized way.

\begin{table}[!ht]
\vspace*{.3cm}
\begin{flushleft}
    {\bf Table~3} \\
    Quantitative effects of GMT reforms.
\end{flushleft} {\small \renewcommand{\arraystretch}{1.3}
\begin{tabular}{llll}
   \hline {\bf (a) Gains from a 15\% GMT with $\phi=0.9$} & Non-haven & Haven & World \\ \hline \hline
   Tax rate after GMT introduction & $20.5\%$ for all & $15.0\%$ for all & ----- \\
   Welfare change in bUSD (\% change) & $95.6$ ($3.8\%$) & $-3.6$ ($-4.1\%$) & $92.0$ ($3.6\%$) \\
   Revenue change in bUSD (\% change) & $153.0$ ($22.6\%$) & $-3.6$ ($-4.1\%$) & $149.3$ ($19.5\%$) \\
   Profit change in bUSD (\% change) & $393.0$ $(41.0\%)$ & $-393.0$ $(-41.0\%)$ & $0$ $(0\%)$ \\
   \hline
  \end{tabular} } \\
    \vspace{0.25cm}

{\small \renewcommand{\arraystretch}{1.3}
\begin{tabular}{llll}
   \hline {\bf (b) Gains from a 16\% GMT with $\phi=0.9$} & Non-haven & Haven & World \\ \hline \hline
   Tax rate after GMT introduction & $20.6\%$ for all & $\begin{cases}
       10.3\% &\text{for small} \\
       16.0\% &\text{for large}
   \end{cases}$ & ----- \\
   Welfare change in bUSD (\% change) & $104.4$ ($4.2\%$) & $-8.8$ ($-9.9\%$) & $95.6$ ($3.7\%$) \\
   Revenue change in bUSD (\% change) & $165.0$ ($24.4\%$) & $-8.8$ ($-9.9\%$) & $156.2$ ($20.3\%$) \\
   Profit change in bUSD (\% change) & $422.3$ $(44.0\%)$ & $-422.3$ $(-44.0\%)$ & $0$ $(0\%)$ \\
   \hline
  \end{tabular} } \\
    \vspace{0.25cm}

 {\small
 \renewcommand{\arraystretch}{1.3}
\begin{tabular}{llll}
   \hline
   {\bf (c) Gains from a 16$\%$ GMT with $\phi=1.0$} & Non-haven & Haven & World \\ \hline \hline
   Tax rate after GMT introduction & $20.8\%$ for all & $16.0\%$ for all & ----- \\
   Welfare change in bUSD (\% change) & $114.5$ ($4.6\%$) & $-8.9$ ($-10.0\%$) & $105.6$ ($4.1\%$) \\
   Revenue change in bUSD (\% change) & $181.1$ ($26.8\%$) & $-8.9$ ($-10.0\%$) & $172.2$ ($22.5\%$) \\
   Profit change in bUSD (\% change) & $461.2$ $(48.1\%)$ & $-461.2$ $(-48.1\%)$ & $0$ $(0\%)$ \\
    \hline
  \end{tabular}  }  \vspace{0.25cm}\\
 {\small {\it Notes}: Panel (a) shows the effects of raising the GMT rate from $t_M=0$ to $0.15$ with partial coverage ($\phi=0.9$); panel (b), from $t_M=0$ to $0.16$ with $\phi=0.9$; and panel (c), from $t_M=0$ to $0.16$ with full coverage ($\phi=1.0$). The tax rates in the initial equilibrium (Regime~0) are $0.186$ in the non-haven and $0.093$ in the haven. The regime-switching rate between Regimes 1 and 2 is $t_M^1 = 15.6\%$. Results are based on the calibrated parameter values described in Table~2. The welfare levels are normalised by the MVPF, $\lambda$. The `haven' here refers to the sum of the $H=40$ haven countries.}
\end{table}

For both $t_M=0.15$ and $0.16$, the non-haven's revenues rise due to weaker tax competition (higher tax rates) and reduced profit shifting, while the havens' revenues fall as the loss of shifted profits outweighs the gains from milder tax competition. The calibrated revenue gain in the non-haven (153--165 billion USD) is close to the combined gains for Europe and the U.S. in~\citet[Table~1]{Barakeetal2022} ($67+58=125$ billion EUR) and worldwide gains (149--156 billion USD) are in the same range as in~\citet[Section~8.2]{Huggeretal2024} (155--192 billion USD). Since falling shifted profits reduce MNEs' private benefits, the non-haven's welfare rises less than their revenues.\footnote{In the havens without MNE headquarters, revenues coincide with welfare.}  


Do the quantitative effects of introducing GMT change if the coverage is extended? The effects of $t_M = 0.15$ with full coverage ($\phi = 1.0$) are unchanged since both the non-haven and the havens set a single tax rate for all MNEs, implying the same outcomes as in Table~3(a). By contrast, introducing $t_M=0.16$ with full coverage yields the non-splitting equilibrium in Regime~1 (Table~3(c)), rather than the splitting one in Regime~2 (Table~3(b)). By eliminating competition for small MNEs, the extended coverage increases revenue by about 16 billion USD and welfare by about 10 billion USD, both in the non-haven and worldwide.\footnote{Comparing Tables~3(b) and 3(c), extending $\phi$ from 0.9 to 1.0 increases the non-haven's revenues by $181.1 - 165.0 = 16.1$ billion USD and their welfare by $114.5-104.4=10.1$ billion USD. Corresponding numbers for the world are $172.2-156.2=16.0$ billion USD and $105.6-95.6=10.0$ billion USD, respectively.} These gains should be compared with the estimated compliance costs for small MNEs, however. 
Applying the results of \cite{Gauletal2022} to our setting, the additional compliance costs are in the range of 10.5 billion USD.\footnote{Using a survey of German MNEs (see \citealp{Beckeretal2012} for details on their characteristics), \cite{Gauletal2022} estimate the average one-time expenses of implementing GMT per firm at 1.6 million EUR. Applying this to the number of small MNEs not covered by GMT in our database ($\approx$6,000; see Table~A.1 in \ref{app: fig.1}), the total compliance costs of newly covered MNEs are 1.6 million EUR $\times$ 6,000 $\times$ 1.1 USD/EUR $\approx$ 10.5 billion USD.} 
Subtracting this amount from the gains of a full-coverage GMT results in net {\em revenue} gains of 5.5 billion USD, but a small net {\em welfare} loss of --0.5 billion USD globally.\footnote{\cite{Cliffordetal2025} also estimate size-dependent compliance costs for German MNEs and similarly find that an increased coverage of the GMT would provide only limited gains for Germany.} 

\subsection{Extensions}\label{subsec: extensions}

In this section we provide two extensions of our basic model and confirm the robustness of our quantitative findings. 

\subsubsection{Decentralised decision-making by tax havens}\label{subsubsec: extensions01}

We modify the basic model by allowing the havens to decide tax policies in a decentralised manner. The decentralised decision-making leads to competition between the havens and thus makes it more likely that they split their tax rates.\footnote{\label{foot: tMb}The new regime-switching rate dividing Regimes~1 and~2, denoted by $t_M^b$, is smaller than that under collective decision-making, i.e., $t_M^b < t_M^1$. Consequently, the parameter range in which Regime~1 emerges is narrower: $[t_M^0, t_M^b) \subset [t_M^0, t_M^1)$.} This is because the commitment to a single GMT rate by a single haven does not induce the non-haven to raise its tax rate as much as the collective commitment by multiple havens. This reduces the gains from committing to a single GMT rate for each haven, and increases the range of GMT rates that lead to a split tax equilibrium. A formal analysis of decentralized decision-making by tax havens is provided in~\ref{app: extensions01}.


Using the calibrated parameter values in Table~2, we reassess the quantitative effects of the GMT reforms in Table~4.\footnote{We can use the same calibrated parameter values as in the collective decision-making case because the analytical expressions for the targeted calibration moments, i.e., the non-haven's rate $t_n^0$ and the share of shifted profits $\sum_h \theta_h^0$ in Regime~0, remain unchanged.} Table~4(a) reports the results for a 15\% GMT rate with partial coverage ($\phi = 0.9$). In contrast to the collective decision-making case in Table~3(a), this reform yields an equilibrium in Regime~2: havens split their tax rates, setting a lower rate for small MNEs (10.1\%) than for large MNEs (15.0\%) (see Proposition~1(ii)).
This result --- a GMT rate of 15\% leading the havens to adopt a split tax system --- aligns with observations that low-tax countries like Ireland have announced to keep their base rate unchanged (in Ireland's case at 12.5\%) while applying additional taxes on large MNEs to comply with the GMT.

As a result of this response by the havens, the revenue and welfare gains from introducing a 15\% GMT rate are smaller than under collective decision-making (Table~4(a) vs. Table~3(a)). Increasing the coverage rate from 90\% to 100\% at the 15\% GMT rate eliminates the split tax system, as shown in Table~4(b), and brings revenue and welfare gains for both the non-haven and the world.\footnote{Comparing Tables~4(a) and~4(b), extending $\phi$ from 0.9 to 1.0 increases the non-haven's revenues by $153.0 - 139.5 = 13.5$ billion USD and their welfare by $95.6-87.2=8.4$ billion USD. Corresponding revenue and welfare gains for the world are $149.3-135.7=13.6$ billion USD and $92.0-83.4=8.6$ billion USD, respectively. These numbers are comparable to those under collective decision-making (see the last paragraph of Section~\ref{subsec: basic}).}

\begin{table}[H]
\vspace*{.3cm}
\begin{flushleft}
{\bf Table~4} \\
GMT reforms under decentralised decision-making by tax havens.
\end{flushleft} \vspace*{-1cm}
\begin{center} {\small \renewcommand{\arraystretch}{1.3}
\begin{tabular}{llll}
   \hline {\bf (a) Gains from a 15\% GMT with $\phi=0.9$} & Non-haven & Haven & World \\ \hline \hline
   Tax rate after GMT introduction & $20.3\%$ for all & $\begin{cases}
       10.1\% &\text{for small} \\
       15.0\% &\text{for large}
   \end{cases}$ & ----- \\
   Welfare change in bUSD (\% change) & $87.2$ ($3.5\%$) & $-3.8$ ($-4.3\%$) & $83.4$ ($3.2\%$) \\
   Revenue change in bUSD (\% change) & $139.5$ ($20.6\%$) & $-3.8$ ($-4.3\%$) & $135.7$ ($17.7\%$) \\
   Profit change in bUSD (\% change) & $359.9$ $(37.5\%)$ & $-359.9$ $(-37.5\%)$ & $0$ $(0\%)$ \\
   \hline
  \end{tabular} } \\
    \vspace{0.5cm}

{\small \renewcommand{\arraystretch}{1.3}
\begin{tabular}{llll}
   \hline {\bf (b) Gains from a 15\% GMT with $\phi=1.0$} & Non-haven & Haven & World \\ \hline \hline
   Tax rate after GMT introduction & $20.6\%$ for all & $15\%$ for all & ----- \\
   Welfare change in bUSD (\% change) & $95.6$ ($3.8\%$) & $-3.6$ ($-4.1\%$) & $92.0$ ($3.6\%$) \\
   Revenue change in bUSD (\% change) & $153.0$ ($22.6\%$) & $-3.6$ ($-4.1\%$) & $149.3$ ($19.5\%$) \\
   Profit change in bUSD (\% change) & $393.0$ $(41.0\%)$ & $-393.0$ $(-41.0\%)$ & $0$ $(0\%)$ \\
   \hline
  \end{tabular} }
  \end{center} 
  {\small {\it Notes}: Panel (a) shows the effects of raising the GMT rate from $t_M=0$ to $0.15$ with partial coverage ($\phi=0.9$) and panel (b), from $t_M=0$ to $0.15$ with full coverage $\phi=1.0$. The tax rates in the initial equilibrium (Regime~0) are $0.186$ in the non-haven and $0.093$ in the haven. The regime-switching rate between Regimes 1 and 2 is $t_M^b = 9.3\%$ (see also footnote~\ref{foot: tMb}). Results are based on the calibrated parameter values described in Table~2. The welfare levels are normalised by the MVPF, $\lambda$. The `haven' here refers to the sum of the $H=40$ haven countries.}
  \vspace{0.5cm}

\end{table}

\subsubsection{Real responses of MNEs' profits to taxes}\label{subsubsec: extensions02}

We extend the basic model to incorporate real responses of MNEs to taxes. To keep the model simple, we assume that the pre-tax profit of an MNE decreases with the non-haven's tax rate according to $\pi = \pi^b/(1 + t_n)$, where the baseline pre-tax profit, $\pi^b \in [\pu, \infty)$, follows a cumulative distribution function. The total pre-tax profits are then
\begin{align*}
\Pi = \int_{\pu}^{\infty} \frac{\pi^b}{1+t_n} dF = \frac{\Pi^b}{1+t_n},
\end{align*}
where $\Pi^b \equiv \int_{\pu}^{\infty} \pi^b dF$ are the total baseline pre-tax profits. This negative effect of taxes on pre-tax profits generally reduces optimal corporate tax rates in both countries, compared to the basic model with pre-tax profits independent of taxes.

To reassess the quantitative effects of the GMT reforms, we recalibrate the parameter values because the non-haven’s tax rate in Regime~0, one of the targeted calibration moments, differs from that in the basic model. Employing the same calibration strategy as before, the new calibration yields an MVPF of $\lambda = 6.0$ as shown in Table~E.1 in \ref{app: extensions02}, compared to $\lambda = 2.1$ in Table~2. When MNEs' profits respond to taxes, the non-haven’s unconstrained tax rate is lower than in the absence of such responses. To match this lower rate with its data counterpart, the MVPF, which captures the relative importance of tax revenues relative to MNEs’ private benefits, must be higher. A higher MVPF raises equilibrium tax rates in both countries. In particular, the regime-switching rate between Regimes~1 and~2 is 15.9\%, slightly higher than 15.6\% in the basic model, above which the havens start splitting their tax rates. 

The quantitative magnitudes of the GMT reforms are reported in Table~E.2. in \ref{app: extensions02} and are similar to those in Table~3. For example, the introduction of a 15\% GMT rate with partial coverage increases welfare in the non-haven and the world by 108.9 and 106.3 billion USD respectively (vs. 95.6 and 92.0 billion USD in Table~3(a)), and reduces the havens' welfare by 2.5 billion USD (vs. 3.6 billion USD in Table~3(a)).


\

\section{Conclusion}\label{sec: conclusion}

This paper has analysed the effects of a global minimum tax (GMT) that is confined to large multinational enterprises (MNEs), thus leaving at least 10\% of the global multinational tax base outside its scope. Using a simple model with profit shifting by heterogeneous MNEs, we have shown that this partial coverage of the GMT gives rise to a sequence of tax competition equilibria between a non-haven and a set of symmetric tax haven countries. In particular, introducing a low GMT rate that still binds the tax havens induces them to commit to a single GMT rate for all MNEs and thus results in higher welfare and tax revenues in both sets of countries. However, a further increase in the GMT rate leads first the tax havens, and then the non-haven, to split their tax rates for large and small MNEs, creating a conflict of interest between the two groups of countries. 

The calibrated version of our model suggests that, upon the introduction of the current GMT rate of 15\%, both the non-haven and the havens set a uniform tax rate (Regime~1), while introducing a 16\% GMT rate leads the havens to split their tax rates and undercut the GMT for small MNEs (Regime~2). We have also checked, under alternative model specifications, that a GMT rate close to 15\% results in a split tax system in the haven countries. Although it is still too early to draw definitive conclusions given the ongoing implementation of the GMT, this finding is consistent with the observation that low-tax countries such as Ireland, Hungary, and Switzerland maintain their statutory corporate tax rates below 15\%, but top up the taxation of large MNEs to the level of the GMT. In terms of welfare and tax revenues, our calibrated model predicts that the non-haven gains while the tax havens lose, resulting in a positive net gain globally.



Our main objective in this paper has been to understand the implications of an incomplete coverage of the GMT. We have done so in a highly stylized model, in order to develop sharp intuitions. With the insights gained, it would be fruitful to extend our basic model to incorporate further important aspects of the GMT and the responses of MNEs. 
One extension would incorporate a splitting response of {\em MNEs}, whereby they restructure into smaller entities to benefit from potentially lower tax rates on small MNEs. Another extension would consider the substance-based income exclusion, i.e., additional tax deductions for MNEs that increase investment and employment. We leave these extensions for further research.

\newpage

{\small \section*{Appendix} 
\setcounter{equation}{0} \global\long\def\theequation{A.\arabic{equation}}

\appendix
\section{Data for Fig.~1}\label{app: fig.1}

We select from the Orbis database MNEs satisfying the following conditions:
\begin{itemize}
\item Global Ultimate Owner with foreign subsidiaries.
The threshold ownership is 50.01\%.
\item C1: MNEs report only consolidated accounts, not unconsolidated accounts.
\item Non-missing revenues (`Operating revenue (Turnover)') and pre-tax profits (`P/L before tax').
\end{itemize}

Column (a) of Table~A.1 reports the pre-tax profits and the total number of MNEs in each year. Column (b) restricts the sample to MNEs subject to the GMT, i.e., those with annual revenues of at least 750 million EUR in two or more of the past four years. Using columns (a) and (b), column (c) shows the share of MNEs subject to the GMT in terms of the pre-tax profits and the number of MNEs.

\

\begin{table}[!ht]
    \begin{flushleft}
    {\bf Table A.1} \\
    MNEs in the Orbis database.\label{table: Orbis} \end{flushleft}
    {\small
    \renewcommand{\arraystretch}{1.3}
    \begin{tabular}{lcccccc}
    \hline
      \multicolumn{1}{c}{}  & \multicolumn{2}{c}{(a) All MNEs} & \multicolumn{2}{c}{(b) MNEs $\ge$ 750mEUR} ~ & \multicolumn{2}{c}{(c) Share $=$ B/A} \\
        Year & Pre-tax profits & Number & Pre-tax profits & Number & Pre-tax profits & Number \\ \hline \hline
        2018 & 2320  & 8656 & 2066  & 2333 & 0.89  & 0.27  \\
        2019 & 2430  & 8416 & 2216  & 2437 & 0.91  & 0.29  \\
        2020 & 1985  & 7920 & 1793  & 2228 & 0.90  & 0.28  \\
        2021 & 3564  & 8803 & 3239  & 2627 & 0.91  & 0.30  \\ \hline
    \end{tabular} }
\end{table} }
\vspace{-0.6cm}
{\small \noindent {\em Source}: Orbis database, own calculations. \\
{\em Note}: Pre-tax profits are in billion EUR.}

\newpage

\section{Proofs of Propositions} \label{app: proofs}

\subsection{Proof of Proposition 1}\label{app: prop01}

Regime~0 can be an equilibrium as long as the haven's unconstrained rate is greater than the GMT rate, $t_M < t_h^0 = t_M^0 \equiv \delta(\lambda-1)/[H(3\lambda-1)]$. A new equilibrium regime emerges when $t_M \ge t_M^0$. Since haven countries are all symmetric, we are concerned with a representative haven country $h$ in what follows.

\vspace{.2cm}

\noindent {\bf Regime 1}: $t_M \in [t_M^0, t_M^1]$, where $t_M^0 = \frac{\delta(\lambda-1)}{H(3\lambda-1)}$ and $t_M^1 \equiv \frac{\delta(\lambda-1)(2\lambda-1)}{H [\lambda(3\lambda-1) -\phi(\lambda-1)^2 ]}$.

When the non-haven sets a single non-GMT rate and the haven sets a single GMT rate, the equilibrium tax rates and payoffs are
\begin{equation}
t_n^1 = \frac{(\lambda-1)(Ht_M +\delta)}{H(2\lambda-1)}, \qquad t_h^1 = t_M. \label{t1}
\end{equation}
\begin{equation*} \label{G1}
\begin{split}
&G_n^1 \equiv G_n(t_n=t_n^1, t_h=t_h^1), \qquad G_h^1 \equiv G_h(t_n=t_n^1, t_h=t_h^1).
\end{split}
\end{equation*}

When the non-haven sets a single non-GMT rate and the haven splits its tax rate, the equilibrium tax rates are
\begin{equation}
t_n^2 = \frac{2(\lambda-1)(\phi H t_M + \delta)}{H[\lambda(3+\phi) -(1+\phi)]}, \qquad t_h^2 = \begin{cases}
    \frac{(\lambda-1)(\phi H t_M + \delta)}{H[\lambda(3+\phi) -(1+\phi)]} &\text{for} \ \pi \in [\pu, \pi_M) \\
    t_M &\text{for} \ \pi \in [\pi_M, \infty)
\end{cases}. \label{t2}
\end{equation}
\begin{equation*}
    \begin{split}
&G_n^2 \equiv G_n(t_n=t_n^2, t_h=t_h^2), \qquad G_h^2 \equiv G_h(t_n=t_n^2, t_h=t_h^2).
\end{split}
\end{equation*}

Given the non-haven's single non-GMT rate, the haven prefers to set a single GMT rate, if
{\small
\begin{eqnarray}
&& G_h^1 -G_h^2 = \label{Ghgap}  \\
&&\frac{\lambda \overbrace{\left[ Ht_M(3\lambda-1)-\delta(\lambda-1) \right]}^{\ge 0} \overbrace{\left[ \delta (\lambda-1)(2\lambda-1) -Ht_M \left\{ \lambda(3\lambda-1)-\phi(\lambda-1)^2 \right\} \right]}^{\ge 0} }{\delta H^2(2\lambda-1)[\lambda(3+\phi) -(1+\phi)]^2} (1-\phi)\po \ge 0 ,
\nonumber
\end{eqnarray} }
where the strict equality holds at $t_M \in \{ t_M^0, t_M^1 \}$. Therefore, the tax-competition equilibrium is that the non-haven sets a single non-GMT rate, and the haven sets a single GMT rate, given by \eqref{t1}. 

\vspace{.2cm}

\noindent {\bf Regime 2}: $t_M \in (t_M^1, t_M^2]$, where $t_M^1 = \frac{\delta(\lambda-1)(2\lambda-1)}{H [\lambda(3\lambda-1) -\phi(\lambda-1)^2 ]}$ and $t_M^2 \equiv \frac{2\delta(\lambda-1)}{H [3\lambda -1 -\phi(\lambda-1)]}$.

Since \eqref{Ghgap} becomes negative for $t_M \in (t_M^1, t_M^2]$, the haven splits its tax rate in response to a non-haven's single non-GMT rate. The non-haven is still not bound by the GMT and continues to choose its unconstrained optimal tax rate. Therefore, the tax-competition equilibrium is that the non-haven sets a single non-GMT rate and the haven splits its tax rate (see \eqref{t2}).

\vspace{.5cm}
\noindent {\bf Regime 3}: $t_M \in (t_M^2, t_M^3]$, where $t_M^2 \equiv \frac{2\delta(\lambda-1)}{H [3\lambda-1 -\phi(\lambda-1)]}$ and $t_M^3 = \frac{2\delta(\lambda-1)(8\lambda-3)}{H(3\lambda-1)(4\lambda-1)}$.

For $t_M > t_M^2$, the non-haven's single-non GMT rate, given by \eqref{t2}, is bound by the GMT and it cannot choose the unconstrained maximising rate in response to the haven's splitting rates. The non-haven thus chooses either a single GMT rate or a splitting tax schedule. In response, the haven must always split its tax rate; otherwise it cannot obtain positive tax revenues.

When the non-haven sets a single GMT rate and the haven splits its tax rate, the equilibrium tax rates and payoffs are
\begin{align} \label{t3}
\begin{split}
t_n^3 = t_M \ \ \text{for} \ \pi \in [\pu, \infty), \qquad
t_h^3 =\begin{cases} t_M/2 &\text{for} \ \pi \in [\pu, \pi_M) \\
t_M &\text{for} \ \pi \in [\pi_M, \infty)
\end{cases}.
\end{split}
\end{align}
\begin{equation*} \label{G3}
\begin{split}
G_n^3 \equiv G_n(t_n=t_n^3, t_h=t_h^3), \qquad G_h^3 \equiv G_h(t_n=t_n^3, t_h=t_h^3).
\end{split}
\end{equation*}

When both the non-haven and the haven split their tax rates, the equilibrium tax rates are
\begin{align} \label{t4}
\begin{split}
t_n^4 = \begin{cases} \frac{2\delta(\lambda-1)}{H(3\lambda-1)} &\text{for} \ \pi \in [\pu, \pi_M) \\
t_M &\text{for} \ \pi \in [\pi_M, \infty)
\end{cases}, \qquad
t_h^4 =\begin{cases} \frac{\delta(\lambda-1)}{H(3\lambda-1)} &\text{for} \ \pi \in [\pu, \pi_M) \\
t_M &\text{for} \ \pi \in [\pi_M, \infty)
\end{cases}.
\end{split}
\end{align}
\begin{equation*} \label{G4}
\begin{split}
G_n^4 \equiv G_n(t_n=t_n^4, t_h=t_h^4), \qquad G_h^4 \equiv G_h(t_n=t_n^4, t_h=t_h^4).
\end{split}
\end{equation*}

Given the haven's splitting rates, the non-haven prefers to set a single GMT rate, if
{\small
\begin{equation} \label{Gngap}
\begin{split}
&G_n^3 -G_n^4 = \frac{ \overbrace{\left[ Ht_M(3\lambda-1)-2\delta(\lambda-1) \right]}^{> 0} \overbrace{\left[ 2\delta(\lambda-1)(8\lambda-3) -Ht_M (3\lambda-1)(4\lambda-1) \right]}^{\ge 0} }{8\delta H(3\lambda-2)^2} (1-\phi)\po \ge 0,
\end{split}
\end{equation} }
which is always fulfilled in Regime~3. The strict equality holds at $t_M = t_M^3$ and it holds that $t_M > t_M^2 > \frac{2\delta(\lambda-1)}{H(3\lambda-1)}$ for any $\lambda > 1/3$. Therefore, the tax-competition equilibrium is that the non-haven sets a single GMT rate, and the haven splits its tax rate, as given in~\eqref{t3}.

\vspace{.5cm}

\noindent {\bf Regime 4}: $t_M \in (t_M^3, 1]$, where $t_M^3 = \frac{2\delta(\lambda-1)(8\lambda-3)}{H(3\lambda-1)(4\lambda-1)}$.

Since \eqref{Gngap} becomes negative for $t_M \in (t_M^3, 1]$, the non-haven splits its tax rate in response to the haven’s split tax rates. Therefore, the tax-competition equilibrium is that both the non-haven and the haven split their tax rates.  \qed 

\vspace{0.5cm}

\subsection{Proof of Proposition 2}\label{app: prop02}

In the following proof, since the haven countries are all symmetric, we are concerned with a representative haven $h$.

\vspace{.2cm}
\noindent {\bf Regime 1}: $t_M \in (t_M^0, t_M^1]$, where $t_M^0 = \frac{\delta(\lambda-1)}{H(3\lambda-1)}$ and $t_M^1 = \frac{\delta(\lambda-1)(2\lambda-1)}{H[\lambda(3\lambda-1)-\phi(\lambda-1)^2]}$.

\vspace{.2cm}
Comparing the welfare levels of countries in Regime 1 with those in Regime 0 gives {\small
\begin{align*}
&G_n^1 - G_n^0 = \frac{\left[ Ht_M(3\lambda-1) -\delta(\lambda-1) \right] \left[ H t_M \lambda^2(3\lambda-1) +\delta(\lambda-1)(7\lambda^2 -8\lambda+2)\right]}{2\delta H(2\lambda-1)(3\lambda-1)^2}\po > 0, \\
\
&G_h^1 - G_h^0 = \frac{\lambda \left[ Ht_M(3\lambda-1) -\delta(\lambda-1) \right] \left[ \delta(\lambda-1)(2\lambda-1) -H t_M \lambda(3\lambda-1)\right]}{\delta H(2\lambda-1)(3\lambda-1)^2}\po \equiv \Delta \\
& \Delta  \
\begin{cases}
    \ge 0 &\text{if} \ t_M \le \frac{\delta(\lambda-1)(2\lambda-1)}{H\lambda(3\lambda-1)} \equiv t_M^{+} \\
    < 0 &\text{if} \ t_M > t_M^{+}
\end{cases}, \\
\
&G_W^1 - G_W^0 = \frac{\left[ Ht_M(3\lambda-1) -\delta(\lambda-1) \right] \left[ \delta(\lambda-1)(11\lambda^2 -10\lambda +2) -H t_M \lambda^2(3\lambda-1)\right]}{2\delta H(2\lambda-1)(3\lambda-1)^2}\po > 0,
\end{align*} }
where the last inequality holds from
\begin{align*}
    t_M \le t_M^1 < \frac{\delta(\lambda-1)(11\lambda^2 -10\lambda +2)}{H \lambda^2(3\lambda-1)}.
\end{align*}
In Regime~1, the welfare effect of the introduction a GMT rate is positive in the non-haven and the world. In the haven, it is positive if $t_M \in [t_M^0, t_M^{+})$ and negative if $t_M \in (t_M^{+}, t_M^1]$.

A marginal increase in $t_M$ leads to
\begin{equation*} 
\frac{\partial G_n^1}{\partial t_M} = \frac{H\lambda^2 t_M +\delta(\lambda-1)^2}{\delta(2\lambda-1)}\po > 0, 
\end{equation*}

\begin{align*}
    &\frac{\partial G_h^1}{\partial t_M} = \frac{\lambda [\delta(\lambda-1) -2H\lambda t_M]}{\delta(2\lambda-1)}\po \begin{cases}
        \ge 0 &\text{if} \ t_M \le \frac{\delta(\lambda-1)}{2H\lambda}\equiv t_M^{++} \\
        < 0 &\text{if} \ t_M > t_M^{++}
    \end{cases}, \\
    &\frac{\partial G_W^1}{\partial t_M} = \frac{\delta(\lambda-1)(2\lambda-1) -H\lambda^2 t_M}{\delta(2\lambda-1)}\po > 0,
\end{align*}
where the last inequality holds because of
\begin{align*}
    t_M \le t_M^1 < \frac{\delta(\lambda-1)(2\lambda -1)}{2H \lambda}.
\end{align*}
In Regime~1, the welfare effect of a marginal increase in the GMT rate is positive in the non-haven and the world. In the haven, it is positive if $t_M \in [t_M^0, \delta(\lambda-1)/(2H\lambda))$ and negative if $t_M \in (\delta(\lambda-1)/(2H\lambda), t_M^1]$.

\vspace{.2cm}

\noindent {\bf Regime 2}: $t_M \in (t_M^1, t_M^2]$, where $t_M^1 = \frac{\delta(\lambda-1)(2\lambda-1)}{H[\lambda(3\lambda-1)-\phi(\lambda-1)^2]}$ and $t_M^2 = \frac{2\delta(\lambda-1)}{H[3\lambda-1-\phi(\lambda-1)]}$.

\vspace{.2cm}

For the non-haven, we see
\begin{align*}
    &\frac{\partial G_n^2}{\partial t_M} \bigg|_{t_M=t_M^1} = \frac{(\lambda-1)[8\lambda^3 -5\lambda^2 -2\lambda +1 +\phi(\lambda-1)^2]}{[ 3\lambda-1 +\phi(\lambda-1) ][ \lambda(3\lambda-1) -\phi(\lambda-1)^2 ]}\phi \po > 0, \\
    &\frac{\partial G_n^2}{\partial t_M} \bigg|_{t_M=t_M^2} = \frac{(\lambda-1)[8\lambda^2-5\lambda +1 +\phi(\lambda-1)]}{[ 3\lambda-1 +\phi(\lambda-1) ][ 3\lambda-1 -\phi(\lambda-1) ]}\phi \po > 0.
\end{align*}
From these results and the fact that $G_n^2$ is a quadratic function of $t_M$, it follows that $G_n^2$ increases with $t_M$ for $t_M \in (t_M^1, t_M^2]$. To prove $G_n^2-G_n^0>0$, it is sufficient to show
\begin{align*}
    &G_n^2 -G_n^0 \big|_{t_M=t_M^1} = \frac{\delta(\lambda-1)^3 [(3\lambda-1)(16\lambda^3-13\lambda^2 +1)-\phi(\lambda-1)^3(8\lambda -3)]}{2H(3\lambda-1)^2[ \lambda(3\lambda-1) -\phi(\lambda-1)^2 ]^2}\phi \po > 0,
\end{align*}
which holds for any $\lambda>1$ and $\phi \in (0,1)$.

For the haven, we get
\begin{align*}
    &\frac{\partial G_h^2}{\partial t_M} \bigg|_{t_M=t_M^1} = -\frac{2\lambda(2\lambda-1)(\lambda-1)^2}{[ 3\lambda-1 +\phi(\lambda-1) ][ \lambda(3\lambda-1) -\phi(\lambda-1)^2 ]}\phi \po < 0, \\
    &\frac{\partial G_h^2}{\partial t_M} \bigg|_{t_M=t_M^2} = -\frac{4\lambda^2(\lambda-1)}{[ 3\lambda-1 +\phi(\lambda-1) ][ 3\lambda-1 -\phi(\lambda-1) ]}\phi \po < 0.
\end{align*}
From these results and the fact that $G_h^2$ is a quadratic function of $t_M$, $G_h^2$ must be decreasing in $t_M$ for $t_M \in (t_M^1, t_M^2]$. To prove $G_h^2-G_h^0>0$, it is sufficient to show
\begin{align*}
    &G_h^2 -G_h^0 \big|_{t_M=t_M^1} = -\frac{\delta\lambda (\lambda-1)^5[3\lambda -1+\phi(\lambda-1)]}{H(3\lambda-1)^2(3\lambda-1)^2[ \lambda(3\lambda-1) -\phi(\lambda-1)^2 ]^2}\phi \po < 0,
\end{align*}  which holds for all $\lambda>1$ and $\phi \in (0,1)$.

For the world, we see
\begin{align*}
    &\frac{\partial G_W^2}{\partial t_M} \bigg|_{t_M=t_M^1} = \frac{(\lambda-1)[4\lambda^3 +\lambda^2 -4\lambda +1 +\phi(\lambda-1)^2]}{[ 3\lambda-1 +\phi(\lambda-1) ][ \lambda(3\lambda-1) -\phi(\lambda-1)^2 ]}\phi \po > 0, \\
    &\frac{\partial G_W^2}{\partial t_M} \bigg|_{t_M=t_M^2} = \frac{(\lambda-1)^2(4\lambda-1 +\phi)}{[ 3\lambda-1 +\phi(\lambda-1) ][ 3\lambda-1 -\phi(\lambda-1) ]}\phi \po > 0.
\end{align*}
Hence $G_W^2$ must be increasing in $t_M$ for $t_M \in (t_M^1, t_M^2]$. To prove $G_W^2-G_W^0>0$, we must show {\small
\begin{align*}
    &G_W^2 -G_W^0 \big|_{t_M=t_M^1} = \frac{\delta(\lambda-1)^3 [(3\lambda-1)(14\lambda^3-9\lambda^2-2\lambda+1)-\phi(\lambda-1)^3(10\lambda -3)]}{2H(3\lambda-1)^2[ \lambda(3\lambda-1) -\phi(\lambda-1)^2 ]^2}\phi \po > 0,
\end{align*} } which holds for all $\lambda>1$ and $\phi \in (0,1)$.

In Regime 2, both the introduction and a marginal increase of the GMT rate raise welfare in the non-haven and the world, but reduce it in the haven.

\vspace{.2cm}
\noindent {\bf Regime 3}: $t_M \in (t_M^2, t_M^3]$, where $t_M^2 = \frac{2\delta(\lambda-1)}{H[3\lambda-1-\phi(\lambda-1)]}$ and $t_M^3 = \frac{2\delta(\lambda-1)(8\lambda-3)}{H(3\lambda-1)(4\lambda-1)}$.

\vspace{.2cm}

For the non-haven, we have {\small
\begin{align*}
    &\frac{\partial G_n^3}{\partial t_M} \bigg|_{t_M=t_M^2} = \frac{(\lambda-1)[2\lambda-1 +\phi(2\lambda+1)]}{2[ 3\lambda-1 -\phi(\lambda-1) ]}\po > 0, \\
    &\frac{\partial G_n^3}{\partial t_M} \bigg|_{t_M=t_M^3} = \frac{(\lambda-1)[\phi(8\lambda-3)-(2\lambda-1)]}{2(3\lambda-1)} \po > 0,
\end{align*} }
so that $G_n^3$ must increase with $t_M$ for $t_M \in (t_M^2, t_M^3]$. To prove $G_n^3-G_n^0>0$, we show that the following must hold for all levels of $\lambda>1$ and $\phi \in (0,1)$:
\begin{align*}
    &G_n^3 -G_n^0 \big|_{t_M=t_M^2} = \frac{\delta(\lambda-1)^2 [(3\lambda-1)(16\lambda^3-13\lambda^2 +3)-\phi(\lambda-1)^2(8\lambda -3)]}{2H(3\lambda-1)^2[ 3\lambda-1 -\phi(\lambda-1) ]^2}\phi \po > 0.
\end{align*}
For the haven, we see
\begin{align*}
    &\frac{\partial G_h^3}{\partial t_M} \bigg|_{t_M=t_M^2} = \frac{\lambda(\lambda-1)}{3\lambda-1 -\phi(\lambda-1)}(1-\phi)\po > 0, \\
    &\frac{\partial G_h^3}{\partial t_M} \bigg|_{t_M=t_M^3} = \frac{\lambda(\lambda-1)(8\lambda-3)}{(3\lambda-1)(4\lambda-1)}(1-\phi) \po > 0,
\end{align*}
which implies that $G_h^3$ decreases with $t_M$ for $t_M \in (t_M^2, t_M^3]$. To prove $G_h^3-G_h^0<0$, we must show that
\begin{align*}
    &G_h^3 -G_h^0 \big|_{t_M=t_M^3} = \frac{\delta\lambda (\lambda-1)^2[8(2\lambda -1)(3\lambda-1) -\phi(8\lambda-3)^2]}{H(3\lambda-1)^2(4\lambda-1)^2}\po < 0.
\end{align*} This inequality holds under our assumptions of $\lambda>1$ and $\phi \in (3/4,1)$, noting that $\phi > \frac{3}{4} > \frac{8(2\lambda-1)(3\lambda-1)}{(8\lambda-3)^2}$.

Considering that both the non-haven and the haven increases their welfare along with a gradual increase in $t_M$, world welfare must also increase as $t_M$ rises. To prove $G_W^3-G_W^0>0$, we show that, for all $\lambda>1$ and $\phi \in (0,1)$, it holds that
\begin{align*}
    &G_W^3 -G_W^0 \big|_{t_M=t_M^2} = \frac{\delta(\lambda-1)^2 [(3\lambda-1)(14\lambda^2-15\lambda+3)-\phi(\lambda-1)^2(10\lambda -3)]}{2H(3\lambda-1)^2[ 3\lambda-1 -\phi(\lambda-1) ]^2}\phi \po > 0.
\end{align*}

In Regime~3, the introduction of a GMT rate increases welfare in the non-haven and the world, while reducing it in the haven. A marginal increase in the GMT rate raises welfare in the non-haven, the haven and the world.

\vspace{.2cm}
\noindent {\bf Regime 4}: $t_M \in (t_M^3, 1]$, where $t_M^3 = \frac{2\delta(\lambda-1)(8\lambda-3)}{H(3\lambda-1)(4\lambda-1)}$.

\vspace{.2cm}

For the non-haven, we see
\begin{align*}
    G_n^4 -G_n^0\big|_{t_M=t_M^3} = \frac{\delta (\lambda-1)^2(8\lambda-3)^2}{2H(3\lambda-1)^2(4\lambda-1)}\phi \po > 0,
\end{align*}
for $\lambda>1$ and $\phi \in (0,1)$. Moreover, the marginal effect of the GMT is always positive: $\partial G_n^4/\partial t_M = (\lambda-1) \phi \po > 0$. With these observations and the fact that $G_n^4-G_n^0$ is linear in $t_M$, we conclude that $G_n^4 -G_n^0 > 0$ for $t_M \in (t_M^3, 1]$.

For the haven, we have
\begin{align*}
    &G_h^4 -G_h^0 = -\frac{\delta \lambda(\lambda-1)^2}{H(3\lambda-1)^2}\phi \po > 0,
\end{align*}
which holds for $\lambda>1$ and $\phi \in (0,1)$. The marginal effect of the GMT is zero: $\partial G_h^4/\partial t_M = 0$ for $t_M \in (t_M^3,1]$. Combining these results leads to $G_W^4 - G_W^0 > 0$ and $\partial G_W^4/\partial t_M > 0$.

In Regime~4, the introduction of a GMT rate increases welfare in the non-haven and the world, while reducing it in the haven. A marginal increase in the GMT rate raises welfare in the non-haven and the world, but has no effect on welfare in the haven.  \qed


\vspace{0.5cm}

\section{Equilibrium welfare and shifted profits}\label{app: figures}

Fig. C.1 shows the welfare levels of the non-haven (panel (a)) and the sum of the havens (panel (b)), and the shifted profits for different GMT rates $t_M$.

The tax-competition equilibrium outcomes under partial GMT coverage (solid curves) are compared with those with full GMT coverage (dashed curves). The equilibrium tax rates under full GMT are as follows.

\vspace{0.3cm}

\noindent {\bf Regime~0}: $t_M \in \left[ 0, t_M^0 \equiv \frac{\delta(\lambda-1)}{H(3\lambda-1)} \right)$. \quad Each of the non-haven and the haven sets its own single non-GMT rate:
    \begin{align*}
        t_n^0 = \frac{2\delta(\lambda-1)}{H(3\lambda-1)} \ \ \text{for} \ \pi \in [\underline{\pi}, \infty), \qquad t_h^0 = \frac{2\delta(\lambda-1)}{H(3\lambda-1)} \ \ \text{for} \ \pi \in [\underline{\pi}, \infty).
    \end{align*}
\noindent {\bf Regime~1f}: $t_M \in \left[ t_M^0 \equiv \frac{\delta(\lambda-1)}{H(3\lambda-1)}, t_M^f\equiv \frac{\delta(\lambda-1)}{H\lambda} \right]$. \quad The non-haven chooses a single non-GMT rate, and the (representative) haven chooses a single GMT rate:
    \begin{align*}
        t_n^{1f} = \frac{(\lambda-1)(Ht_M + \delta)}{H(2\lambda-1)} \ \ \text{for} \ \pi \in [\underline{\pi}, \infty), \qquad t_h^{1f} = t_M \ \ \text{for} \ \pi \in [\underline{\pi}, \infty).
    \end{align*}
\noindent {\bf Regime~2f}: $t_M \in (t_M^f, 1]$. \quad Both the non-haven and the haven choose a single GMT rate:
    \begin{align*}
        t_n^{2f} = t_M \ \ \text{for} \ \pi \in [\underline{\pi}, \infty), \qquad t_h^{2f} = t_M \ \ \text{for} \ \pi \in [\underline{\pi}, \infty).
    \end{align*}
Let $G_n^f$ and $G_h^f$ be the equilibrium payoffs of the non-haven and the havens under full GMT. In Regime~0 ($t_M \in [0, t_M^0)$), the equilibrium tax rates are identical under full and partial GMT coverage, as neither country is constrained by the GMT.

We note $t_M^f > t_M^1 \equiv \frac{\delta(\lambda-1)(2\lambda-1)}{H[\lambda(3\lambda-1)-\phi(\lambda-1)^2]}$, given $\phi \in (0,1)$ and $\lambda>1$. That is, the equilibrium outcomes under partial GMT begins to diverge from those under full GMT once $t_M$ enters the range of Regime~1. This is particularly evident in panels (a) and (c).

\begin{figure}[H]
\begin{center}
    \includegraphics[scale=0.65]{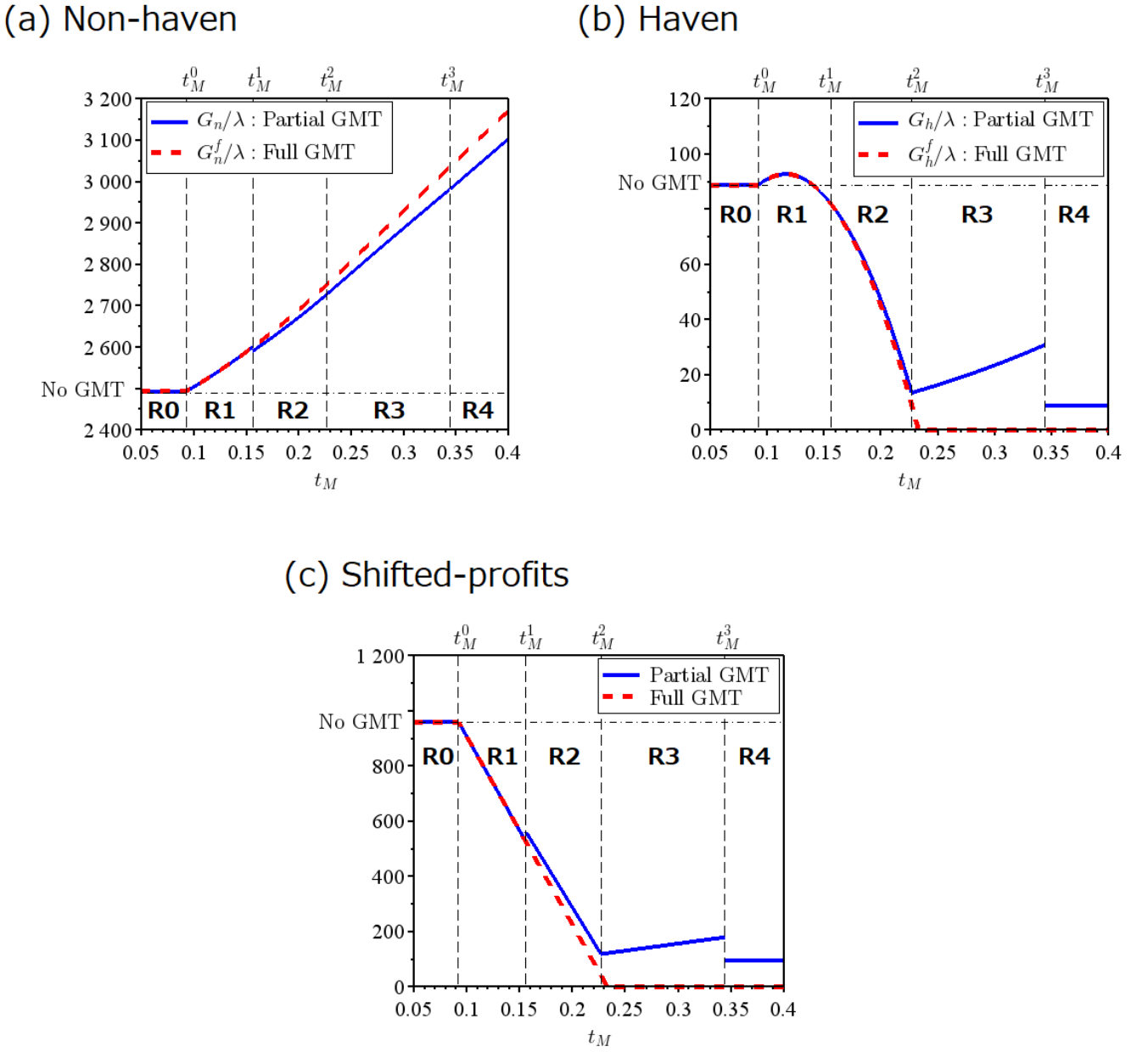}
\end{center}
	\noindent {\bf Fig. C.1.} Equilibrium welfare and shifted profits. \\
   {\small {\it Notes}: The figure shows the equilibrium levels of welfare normalised by the marginal value of public funds (MVPF) $\lambda$ in the non-haven ($G_n/\lambda$) in panel (a) and the sum of the $H=40$ havens ($\sum_{h=1}^H G_h/\lambda$) in (b), and the aggregate shifted profits ($\sum_{h=1}^H \theta_h \Pi$) in (c), for different GMT rates $t_M$. The vertical axis is in billion USD. The solid curves in each panel are the equilibrium values under the partial GMT ($\phi \in (3/4, 1)$); the dashed curves are those under the full GMT ($\phi=1$); horizontal lines are those under no GMT (Regime~0). Results are based on the calibrated parameter values described in Table~2.}
\end{figure}

\section{Welfare effects of gradual GMT reforms}\label{app: gradual}

\subsection{Welfare effects of increasing the GMT rate}\label{app: prop03}

The following proposition characterises the welfare effects of a marginal increase in the GMT rate $t_M$.

\vspace{0.3cm}

\noindent {\bf Proposition D.1. (Increase in the GMT rate $t_M$)}
\vspace{.3cm}

\noindent {\it Consider the tax-competition equilibrium with GMT, as summarized in Proposition~1 in the text. A marginal increase in the GMT rate has the following effects:}
\begin{itemize} {\it 
\item[(i)] welfare in the non-haven increases in all regimes, except at the regime-switching rate $t_M^1$, where it  falls discontinuously.
\item[(ii)] welfare in the (representative) haven increases in Regime 1 as long as $t_M < t_M^{++} \equiv \frac{\delta(\lambda-1)}{2H\lambda}$, then decreases thereafter in Regime 1 and in Regime 2. It increases in Regime 3, falls discontinuously at the regime-switching rate $t_M^3$, and remains unchanged in Regime~4.
\item[(iii)] welfare in the world increases in all regimes, except at the two regime-switching rates $t_M^1$ and $t_M^3$, where it falls discontinuously.}
\end{itemize}

\vspace{0.3cm}

As is true for the introduction of the GMT (Proposition~1), the non-haven generally gains from a gradual increase in the GMT rate $t_M$, as this reduces profit-shifting to the havens in equilibrium. An exception is the regime-switching rate $t_M^1$ where the havens being to split their tax rates and set a discretely lower tax rate for small MNEs. Small MNEs then shift more profits to the havens, hurting the non-haven.

For the havens, the welfare (i.e., tax revenue) effects of an increase in the GMT rate reflect two opposing forces: a negative effect from reduced profit-shifting opportunities and a positive effect from a higher tax rate in the non-haven. These opposing forces generate a non-monotonic welfare response. In Regime~1 ($t_M \in [t_M^0, t_M^1]$), the positive effect dominates initially (when $t_M < t_M^{++}$), but as the GMT rate rises further, the negative effect prevails (when $t_M \ge t_M^{++}$). In Regime~2 ($t_M \in (t_M^1, t_M^2]$), the negative effect dominates throughout. In Regime~3 ($t_M \in (t_M^2, t_M^3]$), the positive effect of a higher tax rate dominates again, as the non-haven commits to a uniform GMT rate in order to raise the havens' tax rates. This regime, where commitment induces effective tax coordination, is similar to the first part of Regime~1 ($t_M \in [t_M^0, t_M^{++})$). It ends at the regime-switching rate $t_M^3$ at which the non-haven starts splitting its tax rate. This leads to a discontinuous fall in the non-haven's tax rate on small MNEs, which must hurt the havens. In Regime~4 ($t_M \in (t_M^3, 1]$), both the non-haven and the havens apply the same GMT rate to large MNEs, preventing them from shifting profits. Consequently, changes in the GMT rate no longer affect the havens' welfare.

The marginal effects of $t_M$ on world welfare are dominated by the effects on the non-haven, and are therefore generally positive. At the two regime-switching rates $t_M^1$ and $t_M^3$, however, one of the countries begins to split its tax rate and discontinuously lowers the tax rate on small MNEs. Therefore, tax competition is tightened and world welfare falls at the margin.

In the following proof, since the haven countries are all symmetric, we are concerned with a representative haven $h$.

\vspace{.3cm}

\noindent {\it Proof:} \quad In \ref{app: prop02}, we already showed the welfare effects of a marginal increase in the GMT rate within each regime. Here we show the welfare effects at the regime switching rates.

At the three regime switching GMT rates, $t_M^1$, $t_M^2$ and $t_M^3$, welfare in the non-haven, the haven, and the world changes as follows.

\vspace{.2cm}
\noindent {\bf Change from Regime 1 to 2 at $t_M^1$}
\begin{align*}
&G_n^2 - G_n^1 \big|_{t_M=t_M^1} =G_W^2 - G_W^1 \big|_{t_M=t_M^1} \\
&= -\frac{\delta(\lambda-1)^3[\lambda(3\lambda-1)(3\lambda-2)-\phi(\lambda-1)^2]}{2H(\lambda-1)[\lambda(3\lambda-1) -\phi(\lambda-1)^2]}(1-\phi)\po < 0, \\
&G_h^2 -G_h^1 \big|_{t_M=t_M^1} = 0.
\end{align*}

\noindent {\bf Change from Regime 2 to 3 at $t_M^2$}
\begin{align*}
&G_n^3 - G_n^2 \big|_{t_M=t_M^2} = G_h^3 -G_h^2 \big|_{t_M=t_M^2} =
G_W^3 - G_W^2 \big|_{t_M=t_M^2} =  0.
\end{align*}

\noindent {\bf Change from Regime 3 to 4 at $t_M^3$}
\begin{align*}
&G_n^4 -G_n^3 \big|_{t_M=t_M^3} = 0, \\
&G_h^4 - G_h^3 \big|_{t_M=t_M^3} =G_W^4 - G_W^3 \big|_{t_M=t_M^3}
= -\frac{8\delta\lambda (\lambda-1)^2(2\lambda-1)}{2H(\lambda-1)[H(3\lambda-1)(4\lambda-1)^2]}(1-\phi)\po < 0.
\end{align*}

This completes the proof of Proposition~C.1. \qed

\vspace{0.5cm}

\subsection{Welfare effects of increasing the GMT coverage rate}\label{app: prop04}

The following proposition characterizes the welfare effects of a marginal increase in the GMT coverage rate $\phi$.

\vspace{.3cm}

\noindent {\bf Proposition D.2. (Increasing in the GMT coverage rate $\phi$)}

\vspace{.2cm}

\noindent {\it Consider the tax-competition equilibrium with GMT, as summarized in Proposition~1 in the text. Except for special cases where a regime switch occurs, a marginal increase in the GMT coverage rate of $\phi \in (0, 1)$ has the following effects:}

\vspace{-.1cm}
\begin{itemize}
\item[(i)] {\it welfare in the non-haven is unaffected in Regime~1 and increases in Regimes~2 to 4.
\item[(ii)] welfare in the (representative) haven is unaffected in Regime~1 and decreases in Regimes~2 to 4.
\item[(iii)] world welfare is unaffected in Regime~1 and increases in Regimes~2 to~4.}
\end{itemize}

\vspace{0.3cm}

In Regime~1, an increase in $\phi$ has no effect on welfare in any country, because all countries levy uniform tax rates. Changing $\phi$ does have effects, however, in the other regimes, where one or both countries split their tax rates. Since the equilibrium tax differential is always greater for small MNEs than for large MNEs, a higher $\phi$, which implies more large MNEs subject to the GMT, reduces aggregate profit shifting. This unambiguously hurts the haven and benefits the non-haven. From the global perspective, a higher GMT coverage is always desirable, as it reduces pressure on tax competition for the profits of small MNEs.

There are two exceptional cases where increasing $\phi$ causes a regime switch.
The first is at $t_M=t_M^1+\epsilon$ with $\epsilon$ being a small positive number, where a rise in $\phi$ pushes Regime~2 back to Regime~1 by weakening the havens' incentive to split their tax rates. This regime switch benefits the non-haven and the world, while it has no effect on the havens. The other is at $t_M=t_M^2+\epsilon$, where a higher $\phi$ triggers a switch from Regime~3 to Regime~2, inducing the non-haven to set a tax rate above the GMT. This latter regime switch, however, leaves welfare in all countries and the world unaffected.

In summary, the welfare effects of an increase in the GMT coverage rate $\phi$ are similar to those of an increase in the GMT rate $t_M$ (Proposition~C.1). A higher $\phi$ has opposing welfare effects in non-haven and haven countries in Regimes~2 to 4, but not in Regime~1. One notable difference from the effects of a higher $t_M$ is that a higher $\phi$ can never harm the non-haven country and the world, even if it induces a regime switch.

In the following proof, since the haven countries are all symmetric, we are concerned with a representative haven $h$.

\vspace{.3cm}

\noindent {\it Proof:} \quad The effects of an increase in the GMT coverage rate, $\phi$, on welfare in the non-haven, the haven, and the world are given as follows.

\vspace{0.2cm}

\noindent {\bf Regime 1}: $t_M \in (t_M^0, t_M^1]$, where $t_M^0 = \frac{\delta(\lambda-1)}{H(3\lambda-1)}$ and $t_M^1 = \frac{\delta(\lambda-1)(2\lambda-1)}{H[\lambda(3\lambda-1)-\phi(\lambda-1)^2]}$.

\vspace{.2cm}
We see
\begin{align*}
&\frac{\partial G_n^1}{\partial \phi} = 0, \quad \frac{\partial G_h^1}{\partial \phi} = 0, \quad \frac{\partial G_W^1}{\partial \phi} = 0.
\end{align*}

\vspace{.2cm}

\noindent {\bf Regime 2}: $t_M \in (t_M^1, t_M^2]$, where $t_M^1 = \frac{\delta(\lambda-1)(2\lambda-1)}{H[\lambda(3\lambda-1)-\phi(\lambda-1)^2]}$ and $t_M^2 = \frac{2\delta(\lambda-1)}{H[3\lambda-1-\phi(\lambda-1)]}$.

\vspace{.2cm}
For the non-haven, we see
\begin{align*}
    &\frac{\partial G_h^2}{\partial \phi} = \frac{\Theta_n [ Ht_M (3\lambda-1) - \delta(\lambda-1)]}{2\delta H[3\lambda-1 +\phi(\lambda-1)]^3}\po, \\
    &\Theta_n \equiv \phi[Ht_M  (\lambda-1)(16\lambda^2 -13\lambda +3)-\delta(\lambda-1)^2] + Ht_M (3\lambda-1)^2 +\delta (\lambda-1)(16\lambda^2 -19\lambda +5).
\end{align*}
The sign of the derivative is determined by that of $\Theta_n$. From $\Theta_n(t_M=t_M^1)>0$, $\Theta_n(t_M=t_M^2)>0$, and the fact that $\Theta_n(t_M)$ is linear in $t_M$, it follows that $\Theta_n>0$.

\vspace{.2cm}
For the haven, we see
\begin{align*}
    &\frac{\partial G_h^2}{\partial \phi} = \frac{\lambda \Theta_h [ Ht_M (3\lambda-1) - \delta(\lambda-1)]}{\delta H[3\lambda-1 +\phi(\lambda-1)]^3}\po, \\
    &\Theta_h \equiv \phi[Ht_M  (\lambda-1)(5\lambda -3)-\delta(\lambda-1)^2] - Ht_M (3\lambda-1)^2 +\delta (\lambda-1)(5\lambda -3).
\end{align*}
The sign of the derivative is determined by that of $\Theta_h$. From $\Theta_h(t_M=t_M^1)<0$, $\Theta_h(t_M=t_M^2)<0$, and the linearity of $\Theta_h(t_M)$ in $t_M$, it follows that $\Theta_h < 0$.

\vspace{.2cm}
For the world, we get
\begin{align*}
    &\frac{\partial G_W^2}{\partial \phi} = \frac{\Theta_W [ Ht_M (3\lambda-1) - \delta(\lambda-1)]}{2\delta H[3\lambda-1 +\phi(\lambda-1)]^3}\po, \\
    \Theta_W &\equiv \phi[Ht_M (\lambda-1)(2\lambda-1)(13\lambda -3)-\delta(\lambda-1)^2(2\lambda+1)] - Ht_M (3\lambda-1)^2(2\lambda-1) \\
    &+\delta (\lambda-1)(26\lambda^2 -25\lambda +5).
\end{align*}
The sign of the derivative is determined by that of $\Theta_W$. From $\Theta_W(t_M=t_M^1)>0$, $\Theta_W(t_M=t_M^2)>0$, and the linearity of  $\Theta_W(t_M)$ in $t_M$, it follows that $\Theta_W > 0$.

\vspace{.2cm}
\noindent {\bf Regime 3}: $t_M \in (t_M^2, t_M^3]$, where $t_M^2 \equiv \frac{2\delta(\lambda-1)}{H [3\lambda-1 -\phi(\lambda-1)]}$ and $t_M^3 = \frac{2\delta(\lambda-1)(8\lambda-3)}{H(3\lambda-1)(4\lambda-1)}$.
\vspace{.2cm}
We see {\small
\begin{align*}
    &\frac{\partial G_n^3}{\partial \phi} = \frac{H(4\lambda-1)t_M^2}{8\delta}\po > 0, \quad \frac{\partial G_h^3}{\partial \phi} = -\frac{H\lambda t_M^2}{8\delta}\po < 0, \quad \frac{\partial G_W^3}{\partial \phi} = \frac{H(2\lambda-1)t_M^2}{4\delta}\po > 0,
\end{align*} }

\vspace{.2cm}

\noindent {\bf Regime 4}: $t_M \in (t_M^3, 1]$, where $t_M^3 = \frac{2\delta(\lambda-1)(8\lambda-3)}{H(3\lambda-1)(4\lambda-1)}$.

\vspace{.2cm}
We get
\begin{align*}
    &\frac{\partial G_n^4}{\partial \phi} = \frac{(\lambda-1)[ 2Ht_M (3\lambda-1)^2 - \delta(\lambda-1)(8\lambda-3) ]}{2H(3\lambda-1)^2}\po > 0, \\
    &\frac{\partial G_h^4}{\partial \phi} = -\frac{\delta\lambda(\lambda-1)^2}{H(3\lambda-1)^2}\po < 0, \\
    &\frac{\partial G_W^4}{\partial \phi} = \frac{(\lambda-1)[ 2Ht_M (3\lambda-1)^2 - \delta(\lambda-1)(10\lambda-3) ]}{2H(3\lambda-1)^2}\po > 0,
\end{align*}
noting that $t_M > t_M^3 > \frac{\delta(\lambda-1)(10\lambda-3)}{2H(3\lambda-1)^2} > \frac{\delta(\lambda-1)(8\lambda-3)}{2H(3\lambda-1)^2}$.

\vspace{.2cm}
This completes the proof of Proposition~C.2. \qed

\

\section{Extensions of the basic model}\label{app: extensions}

\subsection{Decentralised decision-making by tax havens}\label{app: extensions01}

In the text, we assumed that a group of tax-haven countries collectively decide whether to adopt a single GMT rate or to split their tax rate. Here, we consider an alternative, decentralised decision-making process in the first stage of our game, where each tax haven individually decides whether to split, given the decisions of the others. In the following, we focus on $t_M \in [t_M^0, t_M^2]$ and impose two assumptions. First, countries can split their tax rates only if they are subject to the GMT. This assumption reflects the fact that preferential tax regimes for specific tax bases are not permitted under the OECD’s action plan to address base erosion and profit shifting (OECD, 2019). Second, each haven forms the same belief about how the other havens and the non-haven choose their tax rates. This allows us to focus on the problem of a single haven.

Out of the total $H$ tax havens, $H_M$ havens commit to a single GMT rate,
while the remaining $H-H_M$ havens split their tax rates. When the non-haven sets a single non-GMT rate and a haven $h$ sets a single GMT rate, their equilibrium tax rates are respectively
\begin{equation*}
\begin{split}
&t_n(H_M) = \frac{2(\lambda-1)[\phi t_M(H-H_M) +H_M t_M +\delta]}{\phi (\lambda-1)(H-H_M) +\lambda(3H+H_M) -(H+H_M)}, \\
&t_h^1(H_M) = t_M, \label{eq: t(H_M)}
\end{split}
\end{equation*}
Let $G_h^1(H_M)$ denote the equilibrium payoff of the haven $h$ with commitment. We note that as in the text, setting a single non-GMT rate is the non-haven's dominant strategy because if $t_M \in [t_M^0, t_M^2]$, it is never constrained by the GMT: $t_n(H_M) \le t_n(H)=t_n^2 \le t_M$.

If instead the haven $h$ splits its tax rate while the other havens and the non-haven keep their rates unchanged, the number of havens that maintain commitment falls to $H_M-1$. The equilibrium tax rates of the non-haven and the haven $h$ are respectively
\begin{equation}
\begin{split}
&t_n(H_M-1) = \frac{2(\lambda-1)\left[ t_M \left\{ H\phi +(H_M-1)(1-\phi) \right\} +\delta \right]}{(\lambda-1)[\phi (H-H_M+1) + H +H_M -1] +2H\lambda}, \\
&t_h^2(H_M-1) = \begin{cases}
    t_n(H_M-1)/2 &\text{for} \ \pi \in [\underline{\pi}, \pi_M) \\
    t_M &\text{for} \ \pi \in [\pi_M, \infty)
\end{cases}, \label{eq: t(H_M-1)}
\end{split}
\end{equation}
Let $G_h^2(H_M-1)$ be the  equilibrium payoff of the haven $h$ under splitting.

The difference between the haven $h$'s payoffs in the two cases is {\footnotesize
\begin{align*}
    &\Delta G_h^{12}(H_M) \equiv G_h^1(H_M) - G_h^2(H_M-1) \\
    &\ \ = \frac{(1-\phi)\Pi [Ht_M(3\lambda-1) -\delta(\lambda-1)]\Theta(H_M)}{\delta\left[ \phi (\lambda-1)(H-H_M) +\lambda(3H+H_M) -(H+H_M)\right] \left[ (\lambda-1)\left\{ \phi (H-H_M+1) + H +H_M -1\right\} +2H\lambda \right]^2 },
\end{align*} }
where {\small
\begin{align*}
    &\Theta(H_M) \equiv A H_M +B, \\
    &A \equiv (\lambda-1)(1-\phi)\left[ \delta(\lambda-1) -t_M \left\{ H(3\lambda-1) -2(\lambda-1) \right\} \right], \\
    &B \equiv H[\phi(\lambda-1) +3\lambda-1]\left[ \delta(\lambda-1) -t_M \left\{ H(3\lambda-1) -2(\lambda-1) \right\} \right] -2t_M(\lambda-1)^2(1-\phi).
\end{align*} }
If $\Delta G_h^{12}(H_M) \ge 0$ for $H_M \in \{1,2,\dots,H\}$, the haven $h$ always sets a single GMT rate, regardless of how many other havens it believes set a single GMT rate. Under our assumption of symmetric belief formation, all havens end up setting a single GMT rate, and the equilibrium coincides with Regime~1 in the text. Conversely, if $\Delta G_h^{12}(H_M) < 0$ for $H_M \in \{1,2,\dots,H\}$, all havens end up splitting their tax rates, and the equilibrium coincides with Regime~2 in the text.

Since the denominator of $\Delta G_h^{12}(H_M)$ is positive and $Ht_M(3\lambda-1)-\delta(\lambda-1) \ge 0$ for $t_M \in [t_M^0, t_M^2]$, the sign of $\Delta G_h^{12}(H_M)$ is determined
by that of $\Theta(H_M)$. Considering that $\Theta(H_M)$ is linear in $H_M$, the conditions for $\Delta G_h^{12}(H_M) \ge 0$ for $H_M \in \{1,2,\dots,H\}$ reduce to the following two conditions:
\begin{align*}
    &\Theta(1) = A \cdot 1 + B = C - Dt_M \ge 0  &&\Leftrightarrow \ t_M \le t_M^a \equiv C/D, \\
    &\Theta(H) = A H + B = 2(E - Ft_M) \ge 0 &&\Leftrightarrow \ t_M \le t_M^b \equiv E/F,
\end{align*}
where
\begin{align*}
    &C \equiv \delta(\lambda-1)[\phi(\lambda-1)(H-1) +\lambda(3H+1)-(H+1)] > 0, \\
    &D \equiv H\phi(\lambda-1)[\lambda(3H-5)-(H-3)] +H(3\lambda-1)[\lambda(3H-1)-(H-1)] > 0, \\
    &E \equiv \delta H(2\lambda-1)(\lambda-1) > 0, \\
    &F \equiv (6H^2 -4H +1)\lambda^2 -(5H^2 -6H +2)\lambda +(H-1)^2 - \phi(\lambda-1)^2 > 0.
\end{align*}
We can check $t_M^0 < t_M^a < t_M^b < t_M^1$:
\begin{align*}
    &t_M^a - t_M^0 = \frac{2\delta(\lambda-1)^2[\phi(\lambda-1)+3\lambda-1]}{D(3\lambda-1)} > 0, \\
    &t_M^b - t_M^a = \frac{\delta(H-1)(1-\phi)^2(\lambda-1)^4}{DF} > 0, \\
    &t_M^1 - t_M^b = \frac{\delta(H-1)(2\lambda-1)(\lambda-1)^2 [\phi(H+1)(\lambda-1) +\lambda(3H-1) -(H-1)]  }{ F [\lambda(3\lambda-1) -\phi(\lambda-1)^2] } > 0.
\end{align*}
We can also confirm $\Theta(H_M) = AH_M +B$ increases with $H_M$ for $t_M \in [t_M^0, t_M^b]$, or equivalently $A>0$ for $t_M \in [t_M^0, t_M^b]$. Since $A>0$ if $t_M < t_M^c \equiv \delta(\lambda-1)/[H(3\lambda-1)-2(\lambda-1)]$, it suffices to show $t_M^b < t_M^c$:
\begin{align*}
    t_M^c - t_M^b = \frac{\delta(\lambda-1)[\lambda^2 +(H-1)(2\lambda-1) -\phi(\lambda-1)^2]}{F[H(3\lambda-2)-(H-1)]} > 0,
\end{align*}
which is indeed positive.

Using these results, we can thus conclude that (i) if $t_M \in [t_M^0, t_M^a]$, $\Delta G_h^{12}(H_M) \ge 0$ for $H_M \in \{ 1,2,\dots,H\}$; (ii) if $t_M \in (t_M^b, t_M^2]$, $\Delta G_h^{12}(H_M) < 0$ for $H_M \in \{ 1,2,\dots,H\}$; and (iii) if $t_M \in (t_M^a, t_M^b]$, there exists a positive integer $H_M^* \in \{2,\dots,H \}$ such that $\Delta G_h^{12}(H_M) < 0$ for $H_M \in \{1,\dots, H_M^*-1\}$ and $\Delta G_h^{12}(H_M) \ge 0$ for $H_M \in \{H_M^*,\dots,H\}$.

In cases (i) and (ii), the equilibrium is unique. In case (iii), there can be at most three equilibria. First, if each haven believes that fewer than $H_M^*$ havens set a single GMT rate, all havens end up splitting their tax rates, i.e., $H_M=0$. Second, if each haven believes that more than $H_M^*-1$ havens set a single GMT rate, all havens end up doing so, i.e., $H_M=H$. Third, if $H_M$ coincides with $H_M^*$ at which $\Delta G_h^{12}(H_M=H_M^*)=0$, then $H_M=H_M^*$ havens set a single GMT rate while the remaining $H-H_M^*$ havens split their tax rates, consistent with their beliefs.

The discussion is summarised as follows. If $t_M \in [t_M^0, t_M^a]$, the non-haven chooses a single non-GMT rate, and each haven chooses a single GMT rate, as in Regime~1 in the text (Proposition 1(i)). If $t_M \in (t_M^b, t_M^2]$, the non-haven sets a single non-GMT rate, and each haven splits its tax rate and chooses a GMT rate for large MNEs, but a lower rate than the GMT for small MNEs, as in Regime~2 in the text (Proposition 1(ii)). If $t_M \in [t_M^a, t_M^b)$, both Regimes~1 and 2 can be equilibria and either may emerge. If $t_M \in [t_M^a, t_M^b)$ and in addition $H_M$ coincides with $H_M^*$ at which $\Delta G_h^{12}(H_M=H_M^*)=0$, the non-haven sets a single non-GMT rate ($t_n(H_M^*)$ in \eqref{eq: t(H_M-1)}), and there are $H_M^*$ havens setting a single GMT rate $t_M$ and $H-H_M^*$ havens splitting their tax rates ($t_h^2(H_M^*)$ in \eqref{eq: t(H_M-1)}).

A clear difference from Proposition 1 is that the range where only Regime~1 occurs is smaller ($[t_M^0, t_M^a) \subset [t_M^0, t_M^1)$) when havens make their first stage decision in a decentralised manner, implying that they are less likely to commit to a single GMT rate. This is intuitive because the commitment of a haven induces the non-haven to raise its tax rate less than the simultaneous commitment of many havens.

\subsection{Real responses of MNEs' profits to taxes}\label{app: extensions02}

As described in the text, we allow the pre-tax profits of MNEs, $\pi$, to vary in taxes. Specifically, we assume $\pi = \pi^b/(1 + t_n)$, where the baseline pre-tax profit, $\pi^b \in [\pu, \infty)$, follows a cumulative distribution function $F(\pi^b)$. The total pre-tax profits are 
\begin{align*}
\Pi = \int_{\pu}^{\infty} \frac{\pi^b}{1+t_n} dF = \frac{\Pi^b}{1+t_n},
\end{align*}
where $\Pi^b \equiv \int_{\pu}^{\infty} \pi^b dF$ is the total baseline pre-tax profits. We experimented with an alternative specification, $\pi = \pi^b(1 - t_n)$, and found that the quantitative implications of introducing a 15\% GMT rate are similar to those here. 

We adopt the calibration strategy described in Section~\ref{sec: quantification} of the text. Specifically, we calibrate the unconstrained model without GMT (Regime~0) to match key data on international profit shifting. We set the GMT coverage rate to $\phi=0.9$ and the number of haven countries to $H=40$, as in Section~\ref{sec: quantification}. We jointly choose the MVPF $\lambda$ and the profit-shifting cost parameter $\delta$ to minimize the sum of squared distances between the model-implied non-haven tax rate and shifted profits and their empirical counterparts, $(t_n^0 - 0.186)^2+(\sum_h \theta_h^0 - 0.209)^2$, where 
\begin{align*}
    &t_n^0 = \frac{2\left[ \sqrt{H\left[(16\delta + 9H)\lambda^2 - (38\delta - 6H)\lambda + 12\delta + H\right]} -H(3\lambda-1)\right]}{H(8\lambda - 3)}, \qquad t_h^0 = \frac{t_n^0}{2}, \\
    &\theta_h^0 = (t_n^0 -t_h^0)/\delta.
\end{align*}

The calibration results are given in Table~E.1 and the quantitative effects of GMT reforms in Table~E.2. The results are similar to those in Section~\ref{sec: quantification}. As discussed in the text, the only notable difference is the higher calibrated value of the MVPF, $\lambda = 6.0$ (vs. $\lambda = 2.1$ in Table~2). Otherwise, calibrated values and the magnitude of the quantitative findings are similar to those in the basic model. 

\begin{table}[!ht]
\begin{flushleft}
    {\bf Table~E.1} \\
    Calibration of the model with MNEs' pre-tax profits dependent on taxes.
\end{flushleft}
\begin{center}  {\small \renewcommand{\arraystretch}{1.3}
\begin{tabular}{lll}
   \hline
   {\bf Parameter}  &  & Source \\ \hline \hline
   $\phi = 0.9$ & Profit share of MNEs covered by GMT & OECD (2020), Orbis \\
   $H = 40$ & Number of tax havens & T\o rsl\o v et al. (2023) \\
   $\Pi^b =$5,480{\small bUSD} & Baseline pre-tax profits of MNEs & Calibrated \\
    $\lambda=6.0$ & Marginal valuation of public funds & Calibrated \\
   $\delta=17.7$ & Cost of profit shifting & Calibrated \\
   \hline
  \end{tabular} } \\ 
\end{center}
\begin{center}  {\small
\renewcommand{\arraystretch}{1.3}
\begin{tabular}{llll}
   \hline
   {\bf Targeted moment} & Model & Data & Source \\ \hline \hline
   Pre-tax profits of MNEs: {\footnotesize $\Pi=\Pi^b/(1+t_n^0)$} & 4,623bUSD & 4,623bUSD & {\footnotesize T\o rsl\o v et al. (2023), CbCR} \\
   Corp. tax rate of non-haven: {\footnotesize $t_n^0$} & $18.6\%$ & $18.6\%$ &  {\footnotesize T\o rsl\o v et al. (2023)}  \\
   Share of shifted profits: {\footnotesize $\sum_h \theta_h^0$} & $20.9\%$ & $20.9\%$ &  {\footnotesize T\o rsl\o v et al. (2023)}, CbCR \\
   \hline
  \end{tabular} } \\ 
  \end{center}
\begin{center}  {\small
\renewcommand{\arraystretch}{1.3}
\begin{tabular}{llll}
   \hline
   {\bf Non-targeted moment}  & Model & Data & Source \\ \hline \hline 
   Corp. tax rate of haven: {\footnotesize $t_h^0$} & $9.3\%$ & $13.7\%$ & {\footnotesize T\o rsl\o v et al. (2023)} \\
   Non-haven's revenue loss: $t_n^0 \sum_h \theta_h^0 \po$ & $180$bUSD & $247$bUSD & {\footnotesize T\o rsl\o v et al. (2023)} \\
   \hline
  \end{tabular} }
  \end{center}
  {\small {\it Notes}: Data are from 2019. The definitions of non-haven and haven countries follow T\o rsl\o v et al. (2023) and the data is from the authors' website: \url{https://missingprofits.world} (corresponding figures in the spreadsheets of `Table 1' and `Table U1' in the excel file `1975-2019 updated estimates: Tables`). The non-haven countries are $30$ OECD countries, $7$ major developing countries and the rest of the world. There are $40$ haven countries across the world. The corporate tax rates for the representative non-haven and haven countries are calculated as the GDP-weighted averages of their respective effective tax rates. T\o rsl\o v et al. (2023) report the pre-tax profits of foreign-owned MNEs only, which do not include those of domestically-owned MNEs (see their Section 3.1.2 on p.7). Using the CbCR Statistics of the EU Tax Observatory (\url{https://www.taxobservatory.eu/repository/the-cbcr-explorer/}), we compute the pre-tax profit ratio of domestically-owned MNEs to foreign-owned MNEs. With this ratio ($0.785$) and the pre-tax profits of foreign-owned MNEs ($2,590$ billion USD), we obtain the total pre-tax profits of foreign- and domestically-owned MNEs as $(1+0.785)\times 2,590=4,623$ billion USD.}
\end{table}

\begin{table}[!htbp]
\begin{flushleft}
    {\bf Table~E.2} \\
    GMT reforms when MNEs' pre-tax profits depend on taxes.
\end{flushleft} \vspace{-1cm}\begin{center} {\small \renewcommand{\arraystretch}{1.3}
\begin{tabular}{llll}
   \hline {\bf (a) Gains from a 15\% GMT with $\phi=0.9$} & Non-haven & Haven & World \\ \hline \hline
   Tax rate after GMT introduction & $20.7\%$ for all & $15.0\%$ for all & ----- \\
   Welfare change in bUSD (\% change) & $108.9$ ($8.3\%$) & $-2.5$ ($-2.8\%$) & $106.3$ ($7.6\%$) \\
   Revenue change in bUSD (\% change) & $140.7$ ($20.8\%$) & $-2.5$ ($-2.8\%$) & $138.2$ ($18.0\%$) \\
   Profit change in bUSD (\% change) & $386.9$ $(39.9\%)$ & $-386.9$ $(-39.9\%)$ & $0$ $(0\%)$ \\
   \hline
  \end{tabular} } \\
    \vspace{0.5cm}

{\small \renewcommand{\arraystretch}{1.3}
\begin{tabular}{llll}
   \hline {\bf (b) Gains from a 16\% GMT with $\phi=0.9$} & Non-haven & Haven & World \\ \hline \hline
   Tax rate after GMT introduction  & $20.8\%$ for all & $\begin{cases}
       $10.4\%$ &\text{for small} \\
       $16.0\%$ &\text{for large}
   \end{cases}$ & ----- \\
   Welfare change in bUSD (\% change) & $118.1$ ($9.0\%$) & $-7.4$ ($-8.3\%$) & $110.7$ ($7.9\%$) \\
   Revenue change in bUSD (\% change) & $152.0$ ($22.4\%$) & $-7.4$ ($-8.3\%$) & $144.6$ ($18.8\%$) \\
   Profit change in bUSD (\% change) & $416.5$ $(43.0\%)$ & $-416.5$ $(-43.0\%)$ & $0$ $(0\%)$ \\
   \hline
  \end{tabular} } \\
    \vspace{0.5cm}

 {\small
 \renewcommand{\arraystretch}{1.3}
\begin{tabular}{llll}
   \hline
   {\bf (c) Gains from a 16$\%$ GMT with $\phi=1.0$} & Non-haven & Haven & World \\ \hline \hline
   Tax rate after GMT introduction & $21.0\%$ for all & $16.0\%$ for all & ----- \\
   Welfare change in bUSD (\% change) & $129.3$ ($9.9\%$) & $-7.4$ ($-8.3\%$) & $121.9$ ($8.7\%$) \\
   Revenue change in bUSD (\% change) & $166.3$ ($24.5\%$) & $-7.4$ ($-8.3\%$) & $158.9$ ($20.7\%$) \\
   Shifting-profits change in bUSD (\% change) & $453.6$ $(46.8\%)$ & $-453.6$ $(-46.8\%)$ & $0$ $(0\%)$ \\
    \hline
  \end{tabular}  } 
\end{center}
  {\small {\it Notes}: Panel (a) shows the effects of raising the GMT rate from $t_M=0$ to $0.15$ with partial coverage ($\phi=0.9$); panel (b), from $t_M=0$ to $0.16$ with $\phi=0.9$; and panel (c), from $t_M=0$ to $0.17$ with full coverage ($\phi=1.0$). The tax rates in the initial equilibrium (Regime~0) are $0.186$ in the non-haven and $0.093$ in the haven. The regime-switching rate between Regimes 1 and 2 is 15.9\%. Results are based on the calibrated parameter values described in Table~E.1. The welfare levels are normalized by the MVPF, $\lambda$. The `haven' here refers to the sum of the $H=40$ haven countries.}
\end{table}

\pagebreak
\newpage

\begin{spacing}{1.1}

\FloatBarrier 
\bibliographystyle{apalike.bst}
\bibliography{econ_ref_Hayato.bib}


\end{spacing}

\end{document}